\documentclass[journal,transmag]{IEEEtran}

\usepackage{cite}
\usepackage{caption}
\usepackage{subfigure}
\usepackage{graphicx}
\usepackage{amsmath,amssymb,amsfonts}
\usepackage{soul}
\usepackage{multirow}

\ifCLASSINFOpdf
\else
\fi
\begin{document}

\title{\textsc{Droidetec}: Android Malware Detection and Malicious Code Localization through Deep Learning}

\author {
\IEEEauthorblockN{Zhuo Ma\IEEEauthorrefmark{1},
Haoran Ge\IEEEauthorrefmark{1},
Zhuzhu Wang\IEEEauthorrefmark{1},
Yang Liu\IEEEauthorrefmark{1},
Ximeng Liu\IEEEauthorrefmark{2}}
\IEEEauthorblockA{\IEEEauthorrefmark{1}School of Cyber Engineering, Xidian University, Xi’an 710071, China}
\IEEEauthorblockA{\IEEEauthorrefmark{2}College of Mathematics and Computer Science, Fuzhou University, China}
\thanks{This work was supported by the National Natural Science Foundation of China (Grant No. U1804263, U1764263, 61872283, U1708262), the China 111 Project (No. B16037).

Corresponding author: Z. Ma (email: mazhuo@mail.xidian.edu.cn).}}

\IEEEtitleabstractindextext{%
\begin{abstract}
  Android malware detection is a critical step towards building a security credible system. Especially, manual search for the potential malicious code has plagued program analysts for a long time.
  In this paper, we propose Droidetec, a deep learning based method for android malware detection and malicious code localization, to model an application program as a natural language sequence.
  Droidetec adopts a novel feature extraction method to derive behavior sequences from Android applications.
  Based on that, the bi-directional Long Short Term Memory network is utilized for malware detection.
  Each unit in the extracted behavior sequence is inventively represented as a vector, which allows Droidetec to automatically analyze the semantics of sequence segments and eventually find out the malicious code.
  Experiments with 9616 malicious and 11982 benign programs show that Droidetec reaches an accuracy of 97.22\% and an F1-score of 98.21\%.
  In all, Droidetec has a hit rate of 91\% to properly find out malicious code segments.
\end{abstract}

}

\maketitle

\IEEEdisplaynontitleabstractindextext

\section{Introduction}\label{introduction}

\IEEEPARstart{A}{ndroid} systems have gained increasing popularity in smart phones and other mobile intelligent terminals in recent years.
Unpleasantly, the accumulation development and open nature of the platform have also attracted a vast number of malware developers.
According to securelist\cite{Kaspersky2019}, Kaspersky detected 230 million unique malicious and potentially unwanted objects, along with 870 thousand malicious installation packages in the third quarter of 2019.
A representative series is Gustuff\cite{Gustuff2019}, which phished credentials and automate bank transactions for over 100 banking applications and 32 cryptocurrency applications. The devastating influence is difficult to get quick and effective control, as manual analysis is time consuming and it places high demands on the experience of security analysts.

To curb the increasing spread of Android malware, researchers have proposed several solutions for automatic detection.
Most of the existing malware detection methods simply make a judgement about whether an application is malicious or not, and some of the methods attach sensitive permissions and other information to the final result. For security analysts, it lacks direct evidence to support the judgement.
To solve this problem, some detection methods expect to trigger as many conditions as possible by means of UI (User Interface) interactions or automatic testing tools like Monkey\cite{canfora2015detecting}. Such methods are able to catch the abnormal behaviors, but it still takes effort to find malicious code for detailed analysis.
What's worse is that some tricks can easily bypass these methods, such as the deliberately delayed launch of malicious programs, the malicious behaviors triggered by some specific network data packages or sometimes even a simple login interface.
Recently, machine learning has been widely researched and used since there is no need for the user prior knowledge, which makes it possible to automatically detect malware.
Such feature-based methods indeed provide more accurate detection, while it still takes a long time for security analysts to confirm where the malicious code hides.
To the best of our knowledge, there have been only two studies on malicious code localization\cite{Narayanan2018, Li2017}. While both of their analyses are only up to the level of packages, and the localization is likely disturbed by adversarial mixture.
It is currently a challenge to implement accurate and fine-grained malicious code localization.

Inspired by the work of Zhou et al.\cite{Zhou2016AttentionBasedBL} which automatically marks the key points of natural language sentences through the attention mechanism\cite{vaswani2017attention}, we realized that it seems to be a solution for the localization problem.
This work proposes Droidetec, a deep learning based approach for Android malware detection and malicious code localization.
The source code is a collection of program execution logic, and is very much like a natural language with massive jump statements and extra words (produced by developers and code obfuscation).
To this end, Droidetec gives a solution to connect the instruction bytecode segments before and after the jump points, and selectively extracts words as behavior sequence from contextual code. We utilizes an LSTM (Long Short-Term Memory) network for sequence process which allows Droidetec to automatically learn a model of malware patterns.
The major contributions of our work can be summarized as follows:

\textbf{Sequence feature extraction approach.} Instead of searching for independent features, Droidetec implements a depth first invocation traversal on instructions of the bytecode. It analyzes all of the calling relationships to figure out a series of program behaviors.
If an application is running, each possible execution of the application corresponds to a part of the behavior sequence.
Different from a control flow graph, the deep invocation traversal keeps all of the instructions in their original order in the same parent node (parent method).
For each program, a complete behavior sequence is extracted as the feature expression. Through subsequent processing which turns a behavior sequence into a vector sequence, the feature sequences (vector sequences) are finally sent to the deep learning model for training and testing.

\textbf{Automatic malicious code localization.}
Droidetec utilizes a weight distribution strategy to calculate the attention value of the malware behavior sequence, which represents the contribution of scattered methods to the final classification.
Sequence fragments with relatively concentrated contribution values indicate what should be more concerned about in the entire malware code.
Droidetec ultimately grabs these sequence fragments and locates to the corresponding decompiled code, which is regarded as suspected code, along with the corresponding packages, classes and specific method calls.
Malicious code localization may be dispensable for common users, but it provides explanations of the classification and assists analysts in identifying the malicious points in the shortest time.

Note that our method deals with the opcode, the bytecode in the APK (Android application package) files, and the features we analyze are various API sequences. That means we discard other instructions except the $invoke$ instructions that are closely related to program behaviors. Hence whether an application is obfuscated or not makes no difference in our case. However, malicious code dynamically loaded from native shared libraries such as .so (shared object) files do not belong to our analysis scope.
In addition, Droidetec does not detail all code related to malicious behaviors. As a complete malicious behavior from beginning to end may be mixed in among multiple normal code segments, it is a burden for users to scan an excess of code. Instead, we select several code segments which are the most suspicious and provide the methods that are able to use the code.

The rest of the paper is structured as follows.
Section \ref{sec2} presents the architecture along with the methodology of Droidetec.
In Section \ref{sec3}, we evaluate the performance and limitations of Droidetec with application samples.
Section \ref{sec4} reviews related work, and Section \ref{sec5} is a conclusion.

\section{Preliminaries}

\subsection{Semantic element vectorization}
In the program, it is the API invocations that represent specific behaviors, while other operation codes play the role of variable maintenance, logic jump, etc, which are incapable of directly reflecting behavior relevant information. Therefore, we analyze the instructions of an application program and regard each API as a semantic element. However, an API expressed as a character string or a number cannot retain the semantics, as these independent expressions lack the contextual information of sequential behaviors.

The Skip-gram model\cite{mikolov2013distributed} is an efficient method to capture a large number of precise syntactic and semantic word relationships, and eventually learn high-quality distributed vector representations.
Past work on NLP (Natural Language Processing) \cite{Lazaridou2015Combining, DeVine:2014:MSS:2661829.2661974, barbieri2016does} has achieved good results with Skip-Gram and its extended models.

In Droidetec, the Skip-Gram model is utilized to map the API to an $n$-dimensional space, where $n$ is a variable parameter.
In a semantic sequence, each unit is a semantic element and is represented as a vector.
Skip-Gram uses each word ($w_i$) as the input to predict the contextual information($w_{i-k}$, $...$, $w_{i-1}$, $w_{i+1}$, $...$, $w_{i+k}$), and after training we use the weight matrix in hidden layers as a lookup table of word vectors.
In our case, we convert each API to a semantic vector, and the dimension is reduced from the $10^{4}$ (one-hot vector) to $10^{2}$.

\subsection{LSTM network}
The LSTM (Long Short-Term Memory) network \cite{doi:10.1162/neco.1997.9.8.1735, graves2005framewise, Graves2014Hybrid} is an artificial RNN (Recurrent Neural Network) architecture that can process entire sequences of data. Composed of an input gate, an output gate and a forget gate, the LSTM unit can maintain the previous information memory. Through extensive experiments, LSTM networks have demonstrated success in image captioning, machine translation, sentiment classification and other tasks.

In Droidetec, source APK files of Android applications are converted to instruction code and consequently serialization features. For context-based analysis, we leverage the Bi-LSTM (Bidirectional LSTM) network to implement a classification model.
Since the complete behavior sequence of a program is attainable, combining the forward and backward analysis with the bidirectional network model offers a better semantic information transfer.

\subsection{Attention mechanism}
In the serialization features we extract, each portion of the behavior sequence contributes dissimilarly to the final classification result. We desire to find out which parts play the most important role.
A feasible solution is to construct a weight distribution mechanism of serialization features that quantifies each API in the sequence. Areas with high weight distribution have a serious possibility to be where malicious behavior occurs.

In dealing with the regional importance problem, we introduce the attention mechanism.
It implements orient perception and memory access, and actually figures out the consistency of the current input and the target state.
The earliest and most successful utilization of attention mechanism is in computer vision \cite{NIPS2014_5542}, which extracts information from images by adaptively selecting several regions and only specially processing the selected regions.
It has been successfully applied in NLP, especially machine translation \cite{vaswani2017attention}.

In our case, the attention mechanism contributes to discovering key features that may imply malicious behavior.
The calculated attention values expose the suspected segments in the behavior sequence.
Droidetec indicates potentially malicious code by providing specific information, including packages, classes and context, along with the decompiled method code.


\section{Malware Detection Method}\label{sec2}

\subsection{Overview}
\begin{figure*}[htbp]
  \centering
  \includegraphics[width=0.8\textwidth]{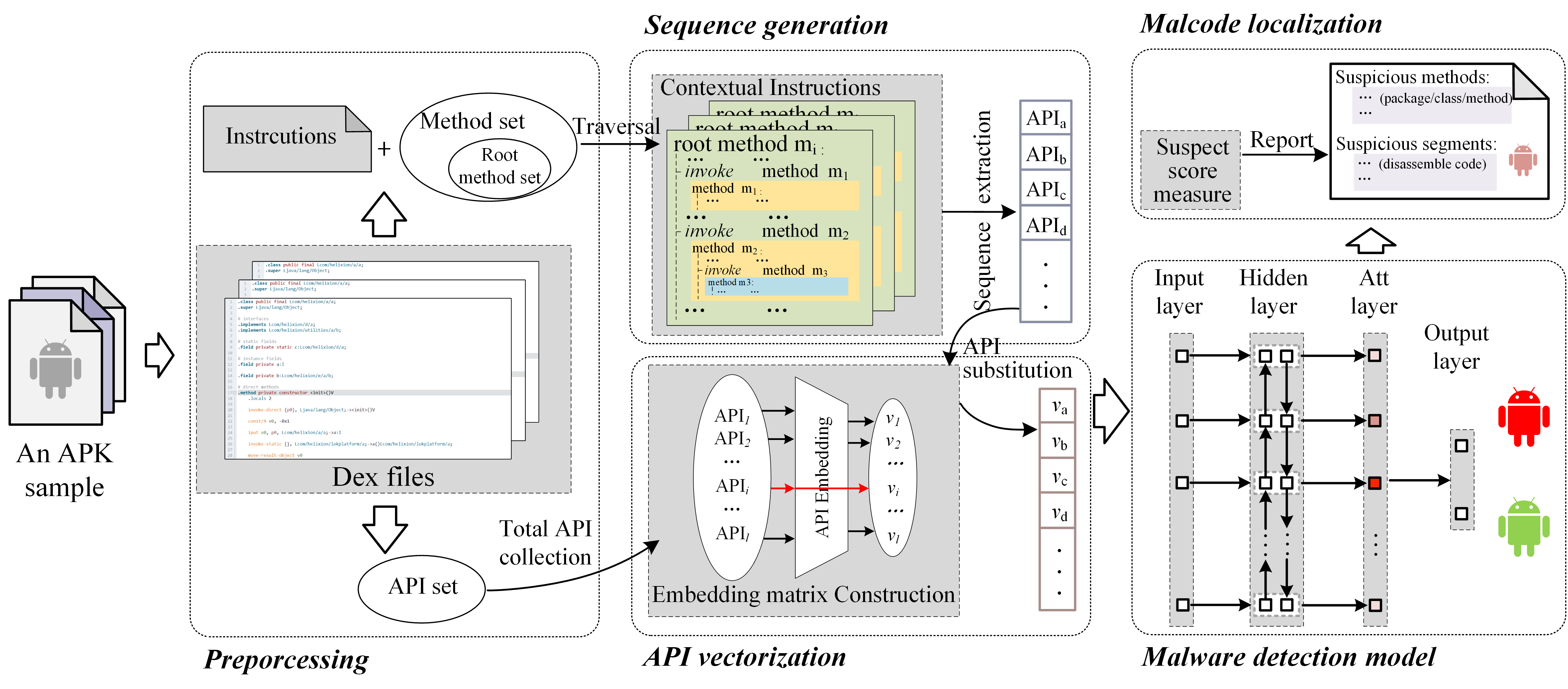}
  \caption{Overview of Droidetec}
  \label{fig:architecture}
\end{figure*}

In this work, we utilize Bi-LSTM network\cite{doi:10.1162/neco.1997.9.8.1735, graves2005framewise} for Android malware detection. As shown in Fig.\ref{fig:architecture}, Droidetec consists of five main stages: the preprocessing, the sequence generation, the API vectorization, the malware detection and malicious code localization.
Training data for Droidetec are behavior sequences from malware samples. We figure out the jump operations in application programs, and statically extract the original behavior sequences that may occur in the runtime of applications. For each original sequence, Droidetec converts it to a vectorial behavior sequence as a piece of training data. Droidetec iteratively trains batches of sequences and maintains the model for malware detection.
Through similar process, the testing stage extracts sequences from extra applications. Once the detection model determines that an application belongs to malware, a deeper analysis could be performed for malicious code localization. The report generated by Droidetec ultimately provides the suspected code segments, including the located packages, classes and methods.
The rest of this section details the five stages.

\subsection{Preprocessing}
We first describe how to deal with an original Android application. In the preprocessing stage, our analysis object is the DEX (Dalvik Executable) file, a compiled Android application code file which contains information on all class files for the entire project. This file format is no trouble to deal with by means of Android reverse engineering tools such as Androguard\cite{desnos2011androguard}.

Decompressed from the APK, the DEX file is then parsed to the corresponding instructions, the method set and the API set. Each instruction consists of opcodes and operation objects, and we only focus on instructions related to the  \textit{``invoke-''} opcodes.
All of the methods defined in the Dex file belong to the method set, which is used in the construction of cross-reference detailed in Section \ref{root_method}. The API set contains all of the invoked APIs, whose names invariably start with \textit{``android/'', ``com/android/internal/util/'', ``dalvik/'', ``java/'', ``javax/'', ``org/apache/'', ``org/json/'', ``org/w3c/dom/'', ``org/xml/sax'', ``org/xmlpull/v1/''} or \textit{``junit/''}. In Droidetec, APIs in package \textit{``java/'', ``javax/''} are outside the scope of analysis as they are huge in number and not related to device behaviors.

\subsection{Sequence generation.}
\label{root_method}
After preprocessing, Droidetec extract the integrated behavior sequence (API sequence) with the instructions and method set.

\textbf{Cross-reference.} For each application, we make statistics to construct the set of all its methods $M$ and combine the instructions corresponding to defined methods. The cross-reference of a method $m_i$ is expressed as two sets, $R_{from}(m_i)$ and $R_{to}(m_i)$.
\begin{gather}
R_{from}(m_i) = \{m|\forall m \in M,\text{if } m_i \text{ directly invokes } m\},\\
R_{to}(m_i) = \{m|\forall m \in M,\text{if } m \text{ directly invokes } m_i\}.
\end{gather}
The method's in-degree $ind(m)$ represents the number of times the method $m$ is invoked. $ind(m)$ can be acquired by calculating the size of $R_{to}(m)$, and the out-degree $outd(m)$ is similarly deduced. The cross-reference reveals the method call relation of an application and lay the foundation for the next step.

\textbf{Root method.} The method call graph of an application can be extremely complicated in most cases. It is almost impossible to find an entry point from which we can grasp the whole behavior code. Some code segments are not executed until the capture of a specific message or signal such as message response functions \textit{onCreate()}, \textit{onStart()} and \textit{onPause()} etc. Thus, we introduce $RM$, the set of root methods, to analyze the whole behavior, which is defined as
\begin{equation}
RM = \{m|\forall m \in M, ind(m)=0\text{ and } outd(m)\ne0\}.
\end{equation}
The in-degree of a root method is 0 while the out-degree is non-zero, which indicates a root method is not explicitly invoked by other methods. A root method is taken as a start point of a series of behavior code.

\begin{figure}[htbp]
  \centering
  \includegraphics[width=0.48\textwidth]{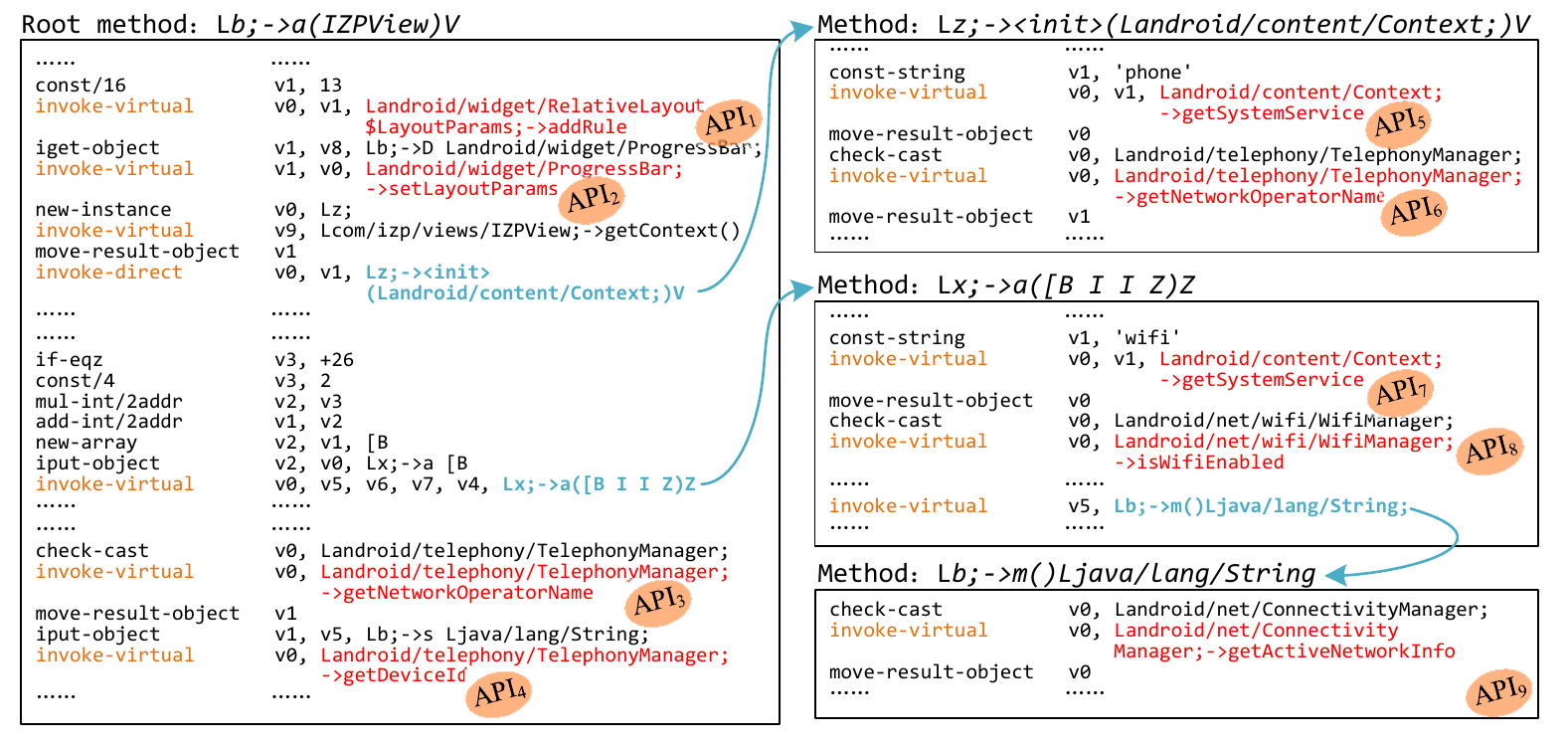}
  \caption{API sequence Extraction}
  \label{fig:API_sequence}
\end{figure}

\textbf{API sequence extraction.} The depth-first invocation traversal is applied to the contextual connection before and after the invocation point. The extraction process starts with the instruction traversal corresponding to root methods in $RM$. Opcodes ranging from 0x6E to 0x72 and 0x74 to 0x78 which represent method invocations are selected for the corresponding operation objects (the method calls). The search process continues recursively to jump to the invoked method and find sequential API calls until the end of the root method. Fig.\ref{fig:API_sequence} depicts an instance of the sequence extraction, and the final sequence arising from root method $b/a()$ is ``$...\to API_1\to API_2\to ...\to API_5\to API_6\to ...\to API_7\to API_8\to ...\to API_9\to ...\to API_3\to API_4\to ...$''.

Droidetec starts from all of the root methods and repeats extraction to generate API sequences, which, to avoid confusion, we call subsequences. These subsequences unite into the complete sequence one by one, which represents the application sample being parsed. Note that APIs are ordered inside each subsequence, while there is no explicit connection between the subsequences, since events are triggered randomly at application runtime.

\subsection{API vectorization}
\label{API_vector}
In the previous process, the complete API sequence is generated, where each API is represented by a serial number that can hardly characterize the difference and correlation between different behaviors.
The numerical representation of various APIs is equivalent to the high-dimensional one-hot vector including a single 1 and other 0s. Assuming that 10000 APIs have been found and API \textit{android/app/Activity;$<$init$>$()} and \textit{android/app/Activity;onCreate()} is respectively labeled with number 1 and 2, the one-hot vectors of the two API can be:
$$
[1\quad \underbrace{0\quad 0\quad \cdots\quad 0}_{9999}] \text{ and }
[0\quad 1\quad \underbrace{0\quad 0\quad \cdots\quad 0}_{9998}].
$$
It can hardly characterize the difference and correlation between different APIs, nor can it describe the context dependence. The appropriate conversion of one-hot vectors is the API distributed representation that improves semantic expression and achieves dimensionality reduction. The converted sequence turns to a vector delivered to the input layer of the detection model. Abundant data parsed from the malicious and benign code are used in multiple rounds of training along with testing.

\textbf{API distributed representation.} Assuming that the total number of API is $l$, each API can be expressed as an $l$-dimensional one-hot vector $a_{si}$, the $i^\text{th}$ API in sequence $s$. We utilize an embedding matrix $W$, just like
$$
 W=
 \left[
 \begin{matrix}
 (w_1)_1 & (w_2)_1 & ... & (w_v)_1\\
 (w_1)_2 & (w_2)_2 & ... & (w_v)_2\\
 \vdots&\vdots&\ddots&\vdots\\
 (w_1)_l & (w_2)_l & ... & (w_v)_l\\
 \end{matrix}
 \right],
$$
to convert the vector and reduce the dimension of $a_{si}$ to $v$.
$W$ is initialized with random values $w$ to represent the weight of dense API representation, and each row uniquely corresponds to an API vector. In this way, a mapping from $a_{si}$ to its distributed representation $v_{si}$ is completed.
\begin{equation}
v_{si} = a_{si} W.
\label{eq:embedding}
\end{equation}

Skip-Gram model is adopted to train the embedding weight matrix $W$ as it performs well in massive corpus. For each extracted API sequence, $v_{si}$ is the input vector to predict the contextual vector $v_{s(i-1)}$, $v_{s(i-2)}$, ..., $v_{s(i-n)}$ and $v_{s(i+1)}$, $v_{s(i+2)}$, ..., $v_{s(i+n)}$. In Skip-Gram, matrix $W$ works as the hidden layer and become a lookup table of distributed presentation API when trained completely.

\subsection{Malware detection}
\begin{figure}[htbp]
  \centering
  \includegraphics[width=0.48\textwidth]{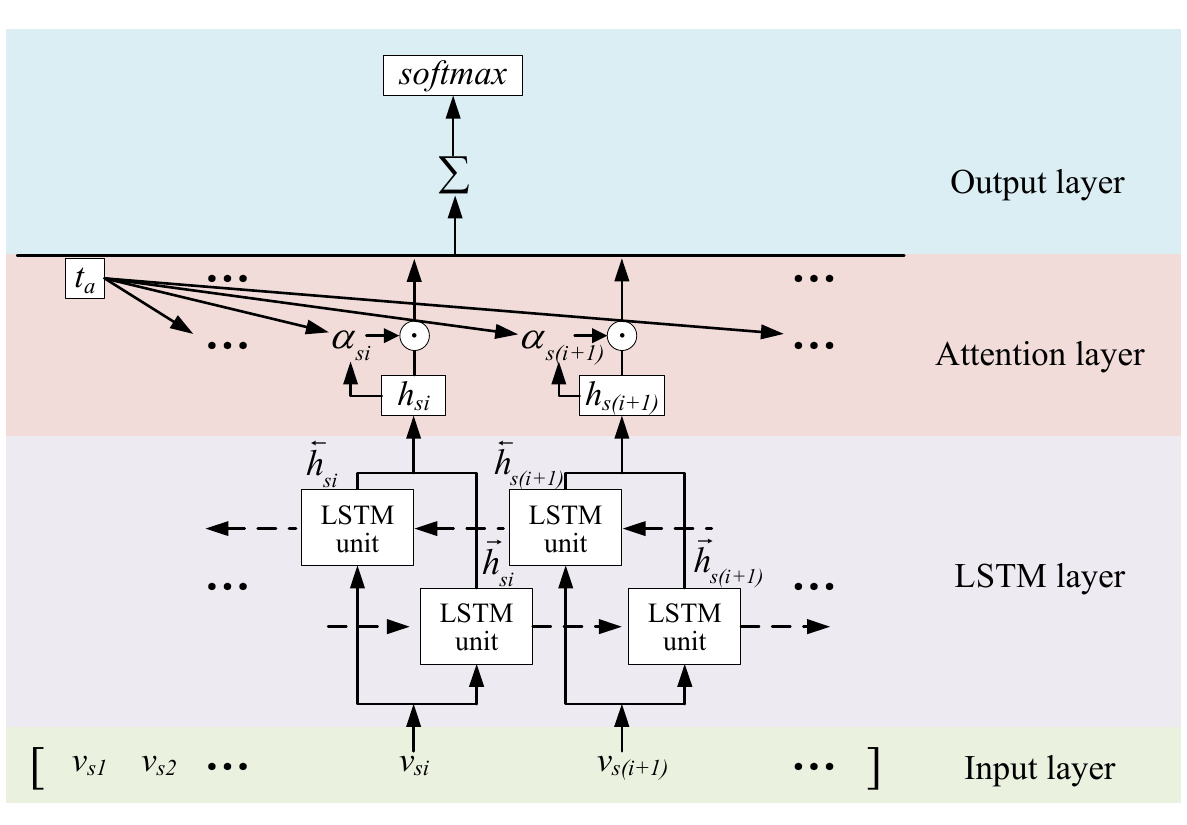}
  \caption{Sequence detection model}
  \label{fig:Att-LSTM}
\end{figure}

The converted sequence $v_{si}$ is used for input data of the sequence detection model.
We adopt the Bi-LSTM network along with an attention layer as the classification model of Droidetec. As shown in Fig.\ref{fig:Att-LSTM}, the model consists of 4 layers: the input layer, the LSTM layer, the attention layer and the output layer.

Each API sequence consisting of distributed representation vectors is passed in the input layer.
The sequences of different applications vary in length, and the size of the input layer is obviously immutable.
We take the maximum value $L$ of all sequence lengths as the size of input layer, and pad the sequence with $v$-dimensional zero vectors at the end if the length is less than $L$.

\begin{gather}
    \overrightarrow{h}_{si} = \overrightarrow{LSTM}(v_{si}), \quad i \text{ from } 1 \text{ to } l,
    \label{eq:forward h}\\
    \overleftarrow{h}_{si} = \overleftarrow{LSTM}(v_{si}), \quad i \text{ from } l \text{ to } 1.
    \label{eq:backward h}
\end{gather}

We then utilize LSTM units to keep the contextual information of a long API vector sequence.
In Eq.\ref{eq:forward h} and \ref{eq:backward h}, forward input and backward input construct the bidirectional LSTM network which leads to a more integral description of API sequences. $l$ represents the length of the API sequence of an application. We utilize $h_{si}$ to summarize the information of a given API vector $v_{si}$ by element-wise summing $\overrightarrow{h}_{si}$ and $\overleftarrow{h}_{si}$.

Not each API contributes equally to the predictive result. We consequently introduce the attention layer to make weight assignment to different $h_{si}$. It figures out which $h_{si}$ we should focus on and the calculated attention values can be assembled to summarize the API sequence. The sequence value of an application is formed by:
\begin{gather}
  t_{si} = \tanh{(h_{si})},\\
  \alpha_{si} = \frac{\exp{(t_a^\mathsf{T} t_{si})}}{\sum_{i}{\exp{(t_a^\mathsf{T} t_{si})}}},
  \label{eq:alpha}\\
  s_s = \sum_{i}{\alpha_{si}h_{si}}.
\end{gather}

The activation function $\tanh$ is used to generate a nonlinear transformation of $h_{si}$. Then the importance weight $\alpha_{si}$ is calculated by a softmax function with $t_{si}$ and $t_a$. In Eq.\ref{eq:alpha}, $t_a$ is a trained parameter vector that represents the state of the API sequence. Ultimately, the complete sequence of the $s^\text{th}$application is expressed as $s_s$, a weighted sum of $h_{si}$ and the corresponding importance weight $\alpha_{si}$.

The final sequence vector $s'$ is a high-level representation of the complete application and is acquired by:
\begin{equation}
  s'_{s} = \tanh{s_s}.
\end{equation}
In the output layer, the sequence vector $s'$ is leveraged in a softmax classifier to make prediction consequently, which is:
\begin{equation}
  p = \mathtt{softmax}(W's'_s+b'),
\end{equation}
where $W'$ and $b'$ are both random initial value for linear regression which represents weight and bias respectively.

\subsection{Malicious code localization}
\label{semantic}
Existing malware detection methods simply make classifications about the malicious degree of applications. Although some of the machine learning based methods have achieved high accuracy, these methods can not distinguish where the threat appears.
In the case of Droidetec, each application predicted as malware is optional for further automatic analysis.
Through malicious code localization, the automatic analysis provides suspected code segments along with relevant details of this malware, which effectively assists security analysts in quick discovery of malicious patterns.

In Eq.\ref{eq:alpha}, the importance weight of each API in the sequence is obtained. The methods defined by the application developer are used as localization units, which can be evaluated by the weighted sum of the invoked APIs. However, it causes deviation that a method can accumulate to a large sum if it invokes a lot of APIs with low weight. A deeper method in the Control Flow Graph (CFG) tends to consist of more APIs, which results in an overvalued weight sum and the excessively wide localization of malicious code. Therefore Droidetec defines the $k$-suspect APIs, $k$ APIs with the highest weights in an API sequence, to reveal which places should we focus on.
The methods directly invoking $k$-suspect APIs at the corresponding positions are the possible malicious methods, and we calculate the suspect scores of these methods with

\begin{gather}
sus(m)=\sum_{i=1}^{k}\alpha_{i}\cdot e_{mi},
\label{eq:suspicious degree1}\\
e_{mi} =
\begin{cases}
1, &\text{if the } i^\text{th} \text{ suspect API exists in } m\\
0, &\text{else.}
\end{cases}
\label{eq:suspicious degree2}
\end{gather}

In Eq.\ref{eq:suspicious degree1} and Eq.\ref{eq:suspicious degree2}, $\alpha_{i}$ represents the weight of the $i^\text{th}$ $k$-suspect API. Droidetec sums up the $\alpha_{i}$ if the corresponding API exists in the suspected method $m$.
The highest $n$ $sus(m)$ are extracted and represented in the report along with the decompiled code.

Note that it is meaningless to compare the $sus(m)$ values in different applications. Inside an application, the sum of the weight $\alpha$ is 1, which means the average of $\alpha$ in a complex application is less than that in the simple. Different $sus(m)$ values can only tell the differences in the suspect degree of methods within the same application.

\section{Evaluation}\label{sec3}
This section focuses on evaluating the classification model and semantic analysis in Droidetec.
All of the experiments have been conducted in a Windows system with an Intel $i5-7400$ CPU.
We first describe the data set we extracted from application samples and our data processing.

\subsection{Dataset}\label{filtering}
\begin{figure}[htbp]
  \centering
  \includegraphics[width=0.4\textwidth]{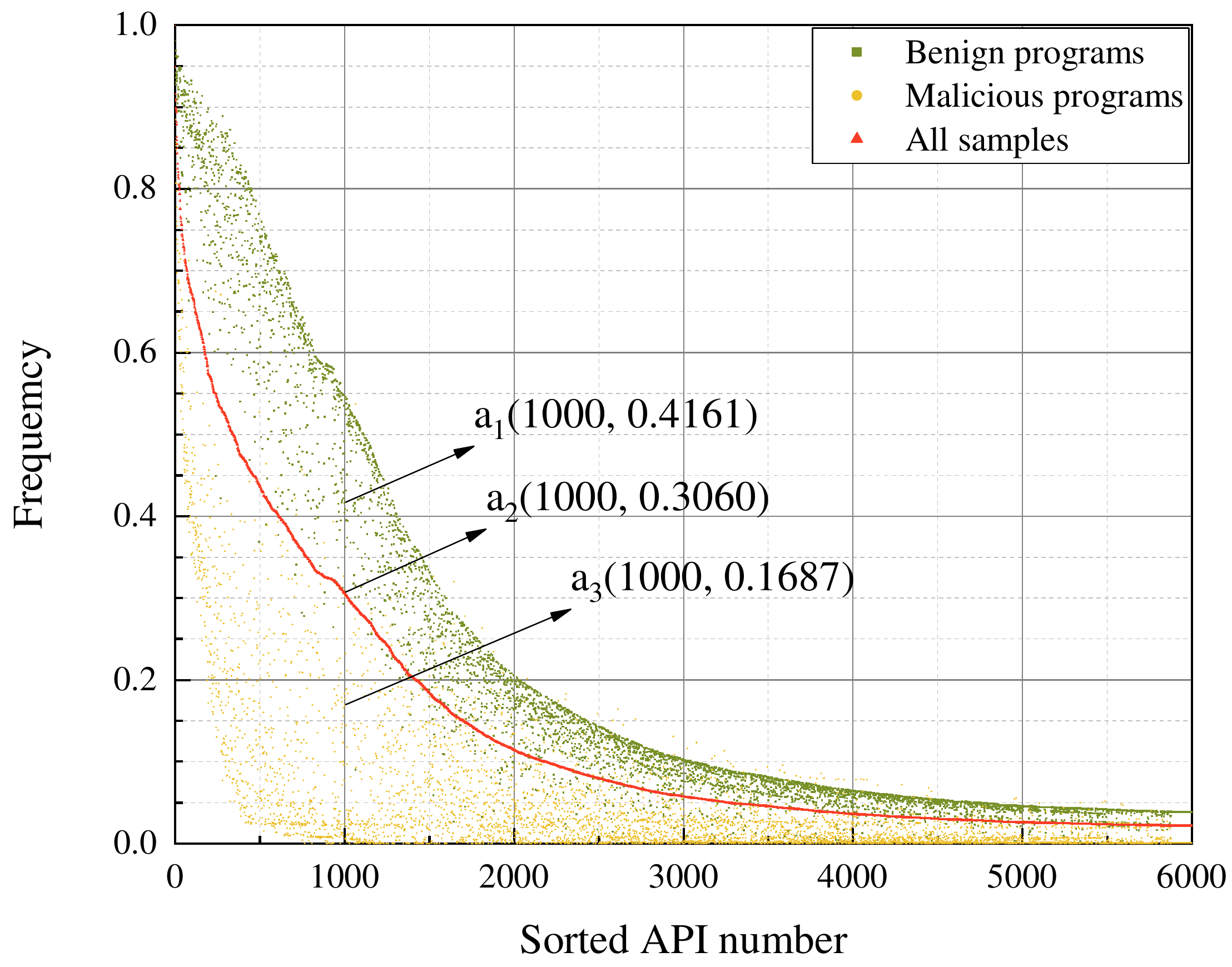}
  \caption{The frequency of each API in applications}
  \label{fig:API_frequency}
\end{figure}

The malware samples we used are from AMD (Android Malware Dataset)\cite{li2017android, wei2017deep}, a carefully-labeled dataset that includes comprehensive profile information of malware, and the benign samples are from Google Play.
A total of 21598 application samples including 9616 malicious and 11982 benign programs cover 65732 different APIs. Before training, we analyzed the usage of each API in the whole program. In Fig.\ref{fig:API_frequency}, the vertical axis represents a certain API's frequency of occurrence while the horizontal axis represents the APIs sorted by its frequency in total applications from high to low. For the sake of presentation, only 6000 APIs are selected as the frequency of remaining APIs tends to be 0. The scatter diagrams respectively represent APIs in benign programs, malicious programs and all samples. For instance, the points $a_1$(1000, 0.4161), $a_2$(1000, 0.3060) and $a_3$(1000, 0.1687) represent the same API \textit{android/os/AsyncTask;onPreExecute()}, which occurs in 6608 samples (30.6\%) including 1622 malicious programs(16.87\%) and 4986 benign programs(41.61\%), and it is the $1000^{th}$ API of all in order of frequency from large to small.

We filtered 43 APIs whose frequency of occurrence are higher than 75\% in malicious, benign and all applications. These are regarded as commonly used APIs that lack in feature tendentiousness, most of which are the basis APIs of a program such as \textit{android/app/Activity;$<$init$>$()} (called in 91.58\% programs, in 94.09\% malware and 88.45\% benign programs, the following are simply expressed as the triad of percentages) and \textit{android/app/Activity;onCreate()} (90.94\%, 92.96\%, 88.44\%) or fairly common APIs such as \textit{android/ content/Intent;$<$init$>$()} (94.89\%, 98.06\%, 90.94\%) and \textit{android/content/Context;getSystemService()} (90.8\%, 94.74\%, 85.88\%).

The data set for training and test includes 21598 API sequences consisting of the remaining 65689 APIs.

\subsection{Detection evaluation}

\subsubsection{Detection performance}
We first work over the effects of API vectors on detection performance.

\begin{figure*}[htbp]
  \centering
  \subfigure[Detection rate]{
  \includegraphics[width=0.3\textwidth]{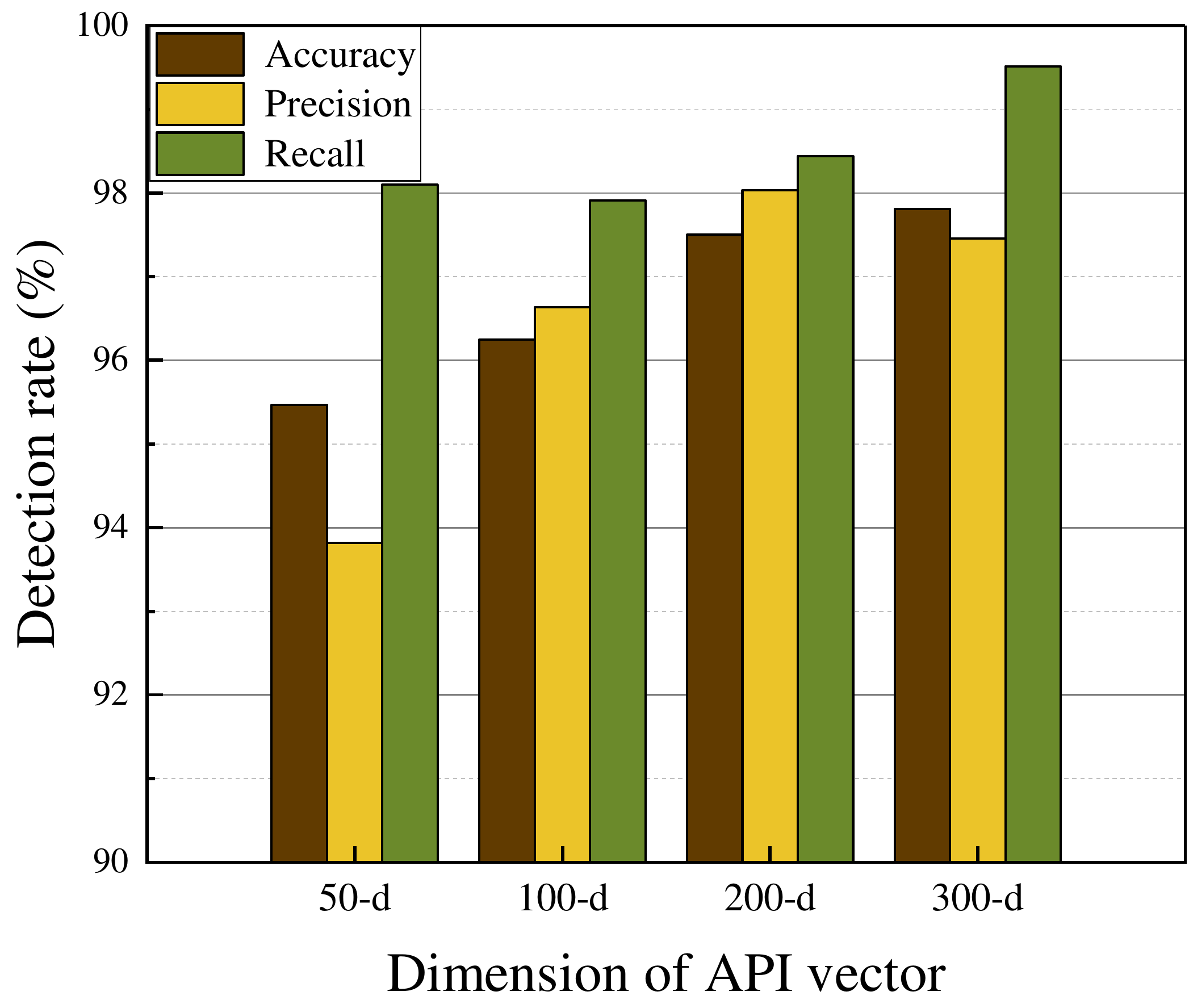}
  \label{fig:API_dimension1}
  }
  \hspace{1.4in}
  \subfigure[F1-score and verification time]{
  \label{fig:API_dimension2}
  \includegraphics[width=0.35\textwidth]{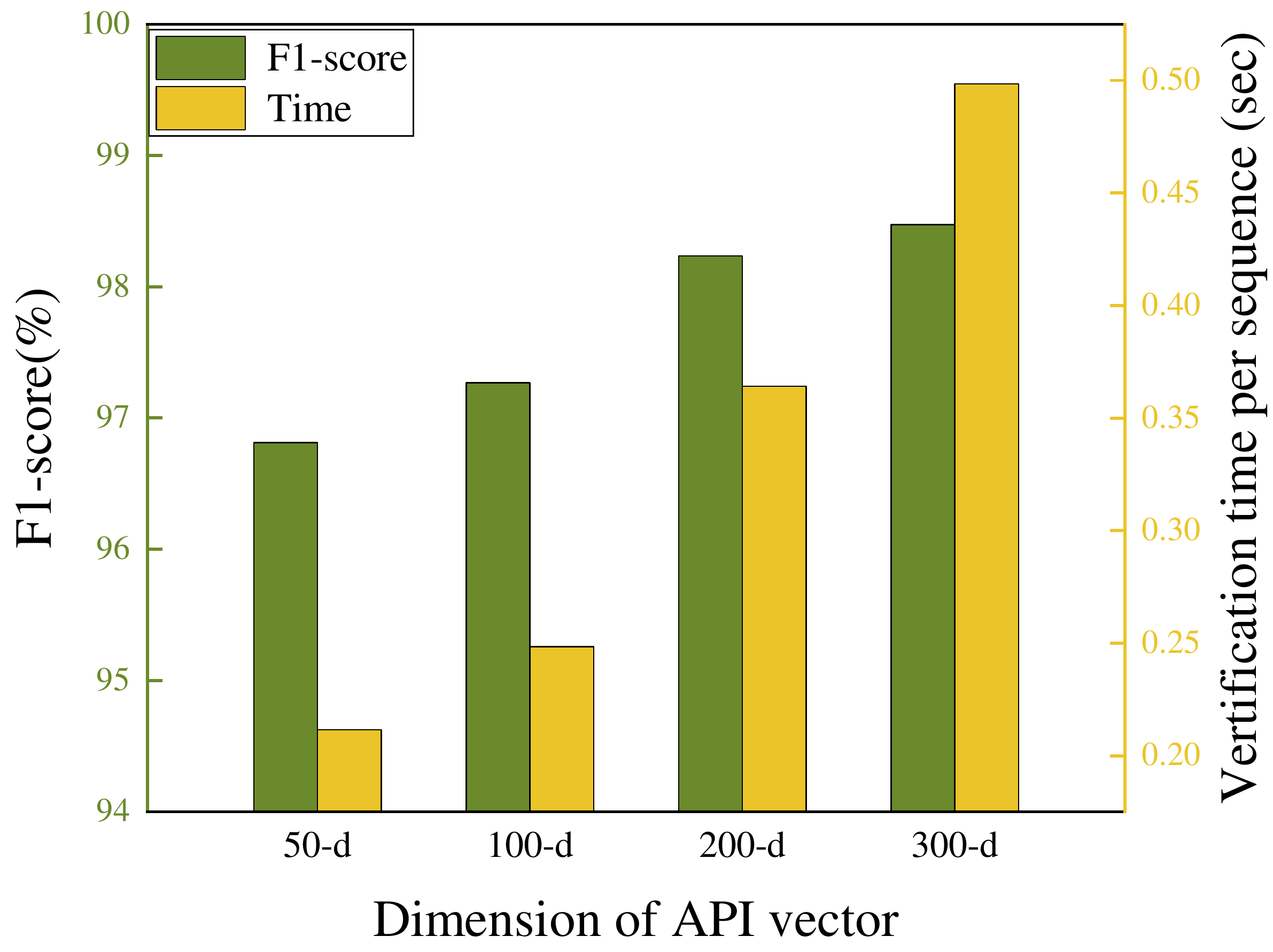}
  }
  \caption{Detection performance of Droidetec}
  \label{fig:detection_performance}
\end{figure*}

In Section\ref{API_vector}, we have described the distributed representation of APIs. Each API is expressed as a vector $v_{si}$ whose dimension $v$ determines the ability of semantic representation. Here we figure out how the dimension $v$ affects the detection performance.
In Fig.\ref{fig:API_dimension1}, accuracy, precision and recall are used to describe the detection performance in different dimensions. In general, Droidetec performs better with the increase of dimension $v$. The detection accuracy increases with the dimension, and the precision and recall rates are generally on the rise. Especially, the recall has remained at around or more than 98\%, which means almost all of the malicious samples can be identified in our test.

However, simply increasing the dimension leads to a computational burden. To better display the effect of dimension changes, we synthetically considered the F1-score and the verification time. The F1-score is the harmonic mean of the precision and recall and expressed by:

\begin{equation}
\textit{F1-score}=\frac{2\times \textit{P}\times\textit{R}}{\textit{P}+\textit{R}},
\end{equation}
where the \textit{P} and \textit{R} respectively represent the precision and recall. It is vividly depicted in Fig.\ref{fig:API_dimension2} that both the F1-score and time consumption increase with the dimension. When the dimension $v$ changes from 200 to 300, the F1-score increases by 0.238\% while the verification time has a 36.8\% increase, from 0.364$s$ to 0.498$s$ per sequence.
In the case of Droidetec, we leveraged the 200-dimension vector to express each distributed API vector that guaranteed efficiency and detection rate (97.2\% accuracy and 98.2\% F1-score).

Besides, we considered the detection performance of Droidetec with various malware families. It is significant for the model to maintain stable and efficient detection of each family.

\begin{figure}
  \centering
  \includegraphics[width=0.4\textwidth]{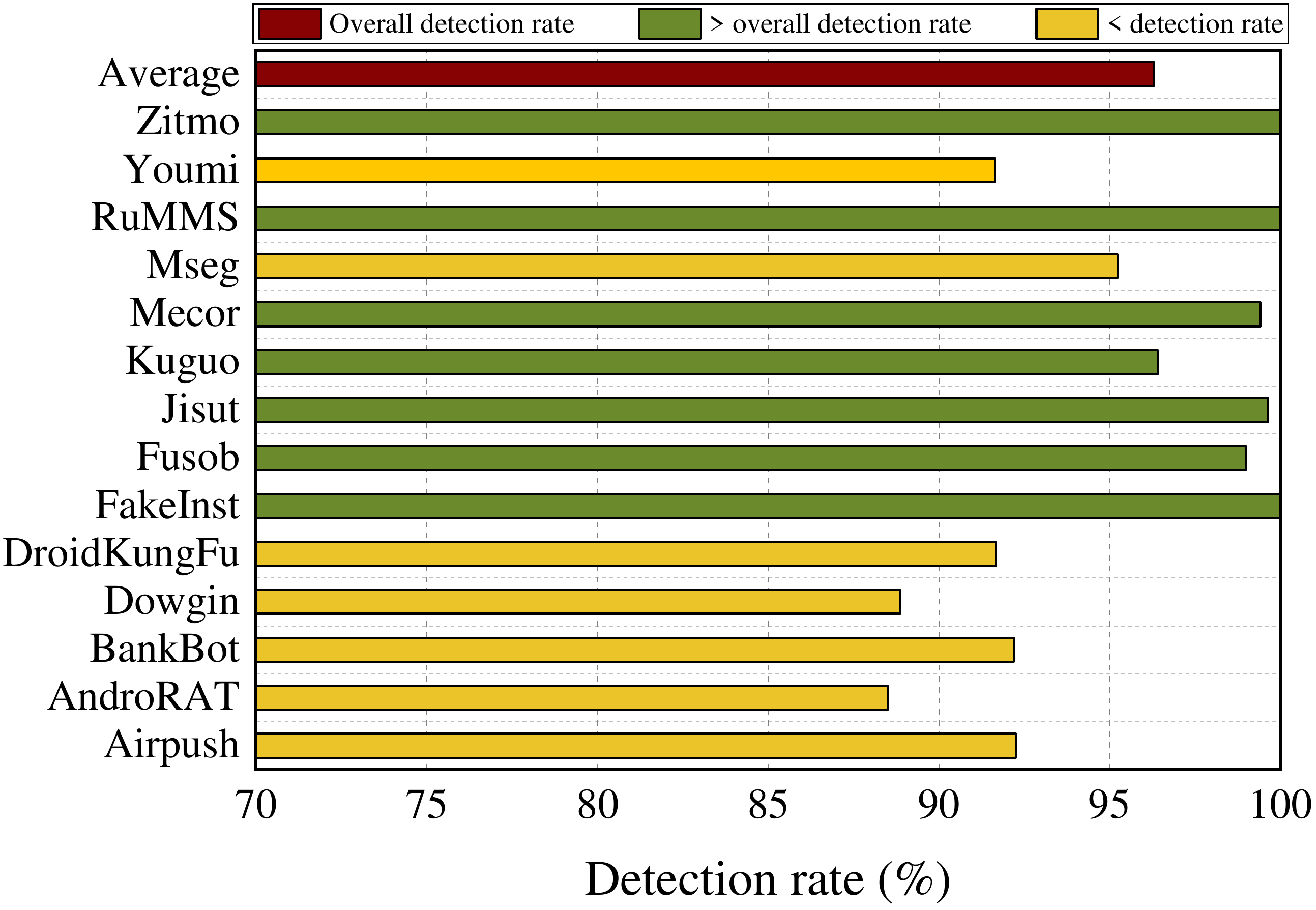}
  \caption{Detection rate in malware families}
  \label{fig:family_detection_rate}
\end{figure}

We took 5850 malware from 14 known families and evaluated the detection performance in Fig.\ref{fig:family_detection_rate}. The overall detection rate of these malicious programs reaches 96.3\%. Especially, Droidetec effectively detects malware from certain families, including \textit{FakeInst}, \textit{Fusob} and \textit{Jisut} etc, with a detection rate of almost 100\%. In two families, \textit{Dowgin} with 270 samples and \textit{AndroRAT} with 40 samples, whose detection rates are respectively 87.8\% and 87.5\%, the insufficient sample training leads to unsatisfactory results. In all, Droidetec generally maintains stable detection capabilities among various malware families.

\subsubsection{Comparison with existing methods}
In Tab.\ref{tab:comparison with existing methods}, we compare the performance of Droidetec with the Droid-Sec \cite{yuan2014droid}, the method proposed by Zhao et al.\cite{8465540} and two methods based on API usage and requested permissions using SVM. Zhao's method uses an ensemble model based on decision tree and k nearest neighbor classifier to analyze sensitive API calls. In our experiment, it reaches a 91.92\% accuracy rate and 90.5\% F1-score with a 9.48\% false positive rate. Droid-Sec was reproduced with 173 features (100 permissions, 62 sensitive API functions and 13 dynamic actions) based on deep belief networks.

Droidetec is superior to both of the two methods in detection accuracy and error rate. Although Droid-Sec utilizes droidbox to capture dynamic behaviors, Droidetec has an F1-score 2.87\% higher and a false positive rate of 1.58\% lower. It reflects that our API sequence based method partially achieves the effect of dynamic detection.

\begin{table*}
  \centering
  \caption{Detection performance comparison}
  \label{tab:comparison with existing methods}
  \begin{tabular}{p{3cm}|p{2cm}|p{2cm}|p{2cm}}
    \hline
    \textbf{Method}& \textbf{Accuracy}& \textbf{F1-score}& \textbf{FPR}\\
    \hline
    Droidetec& 97.22\%& 98.21\%& 2.11\%\\
    \hline
    Droid-Sec\cite{yuan2014droid}& 96.50\%& 95.34\%& 3.69\%\\
    \hline
    Zhao's\cite{8465540}& 91.92\%& 90.50\%& 9.48\%\\
    \hline
    API usage& 83.25\%& 81.56\%& 16.71\%\\
    \hline
    Permissions& 73.11\%& 70.71\%& 26.72\%\\
    \hline
  \end{tabular}
\end{table*}

\subsubsection{Comparison with Android malware scanners}
We then compare the performance of Droidetec with other Android malware scanners.
6 popular scanners are tested with the same data set.

\begin{figure}[htbp]
  \centering
  \includegraphics[width=0.3\textwidth]{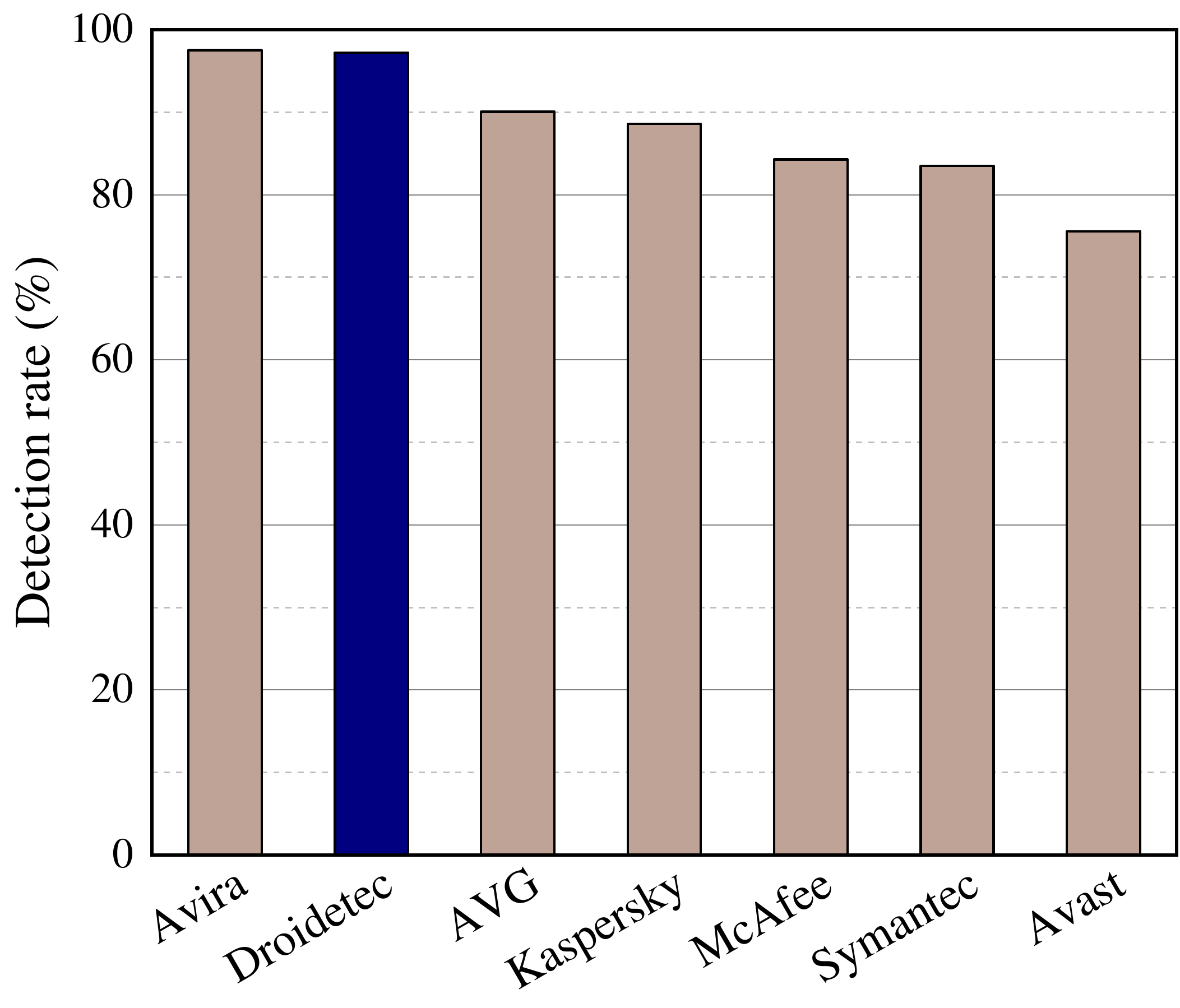}
  \caption{Comparison with Android malware scanners}
  \label{fig:comparison_with_malware_scanners}
\end{figure}

Fig. \ref{fig:comparison_with_malware_scanners} shows the detection rate of Droidetec and mainstream scanners (Avira, AVG, Kaspersky, McAfee, Symantec and Avast). Both Avira and Droirtect demonstrate similar detection capabilities and provide stable detection of a variety of malware families. Other scanners, despite their good performance in most of the samples, have a low detection rate in several families. Especially, some of these scanners present a fairly high tolerance for adware.

\subsection{Semantic analysis evaluation}
Section\ref{semantic} has illuminated that Droidetec scans malicious code based on an attention-based semantic analysis mechanism. This section focuses on the evaluation of semantic analysis between malicious and benign programs and among various malware families.

\subsubsection{Semantic difference in malicious and benign programs}
It is known that different APIs contribute variously to the detection result, while the same API can be important in some programs while less important in others. To verify that Droidetec can distinguish the semantic differences among various program instructions, we depict the distribution of the API attention.

\begin{figure}[htbp]
  \centering
  \subfigure[Benign samples]{
  \includegraphics[width=0.21\textwidth]{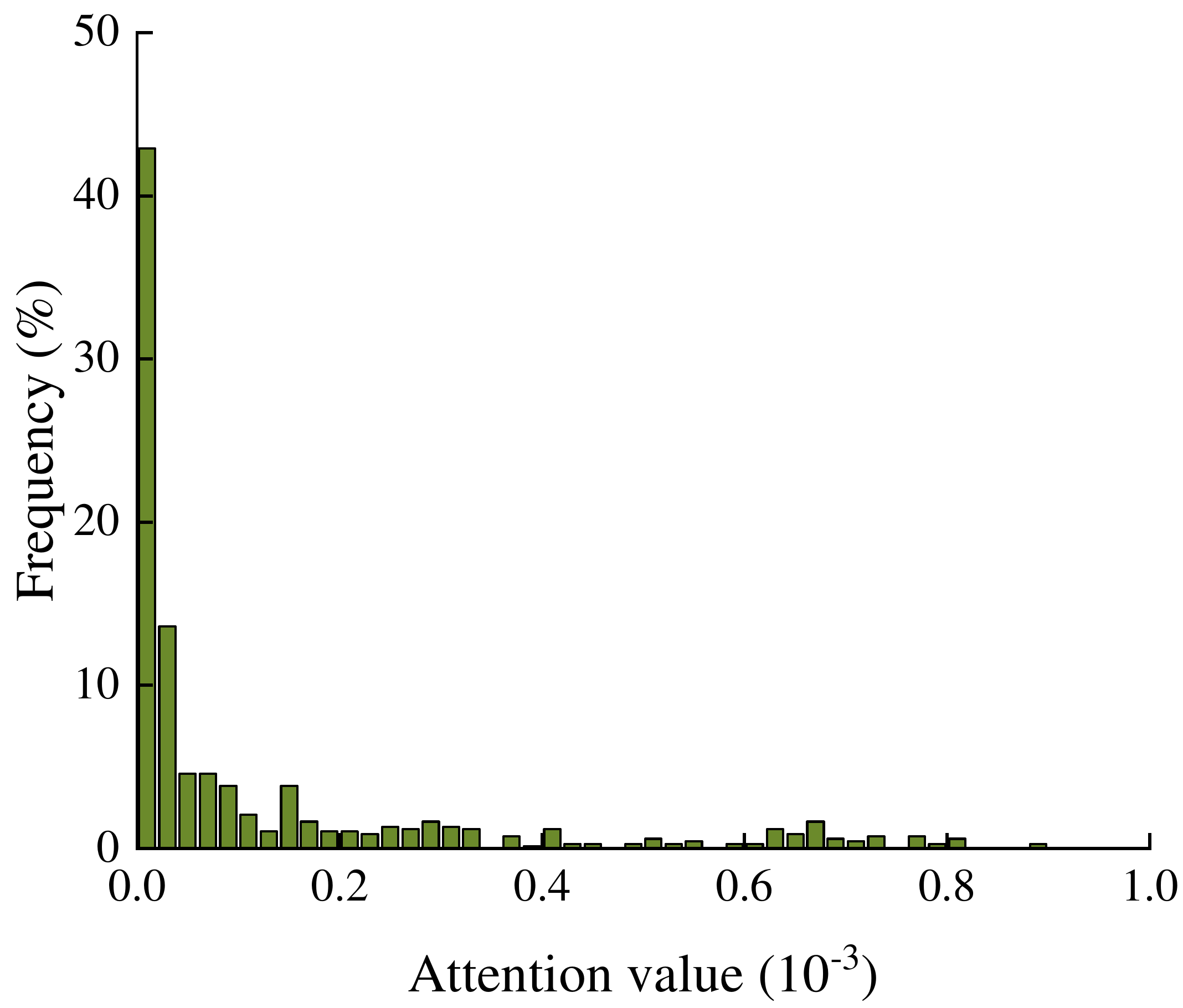}
  }
  \quad
  \subfigure[Malicious samples]{
  \includegraphics[width=0.21\textwidth]{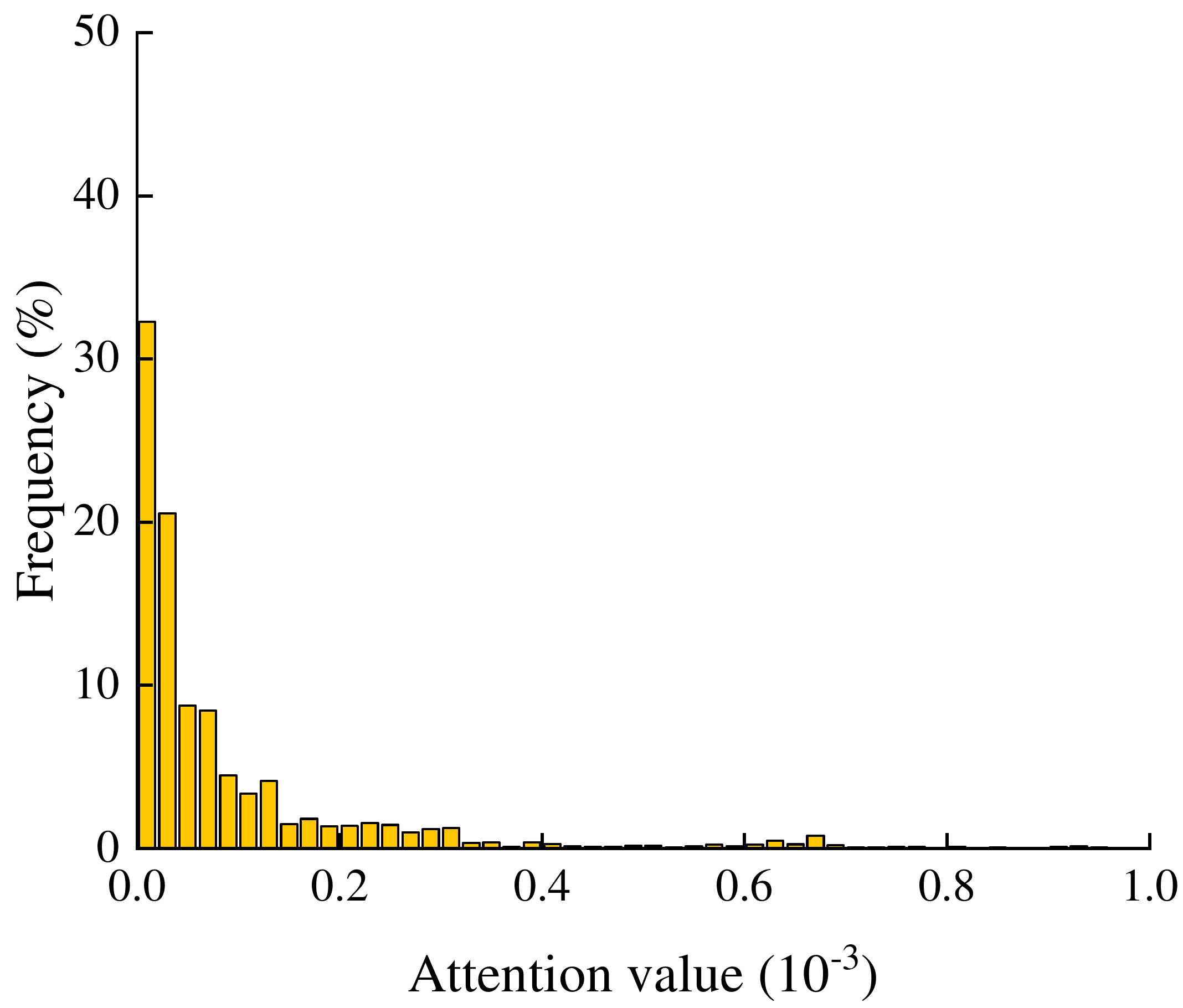}
  }
  \quad
  \caption{API attention distribution of \textit{android/app/Activity; onCreate()}}
  \label{fig:API_distribution1}
\end{figure}

We first studied the distribution of attention values in basis APIs without the previous filtering work in Section\ref{filtering}.
Fig.\ref{fig:API_distribution1} shows the API attention distribution of \textit{android/app/Activity; onCreate()}(abbreviated as \textit{onCreate()}) respectively in benign and malicious samples. As shown, the highest frequency interval is (0, 2$\times{\text{10}}^{\text{-5}}$], and more than 50\% of its attention values of fall within the interval (0, 4$\times{\text{10}}^{\text{-5}}$] in both samples. The consistency in size and distribution indicates that these fairly common APIs, just like \textit{onCreate()} in the example, are given the same low level of attention by Droidetec in both benign and malicious programs. It also shows that the previous filtering work has little impact on subsequent analysis.

\begin{figure*}[htbp]
  \centering
  \subfigure[Benign samples]{
  \includegraphics[width=0.26\textwidth]{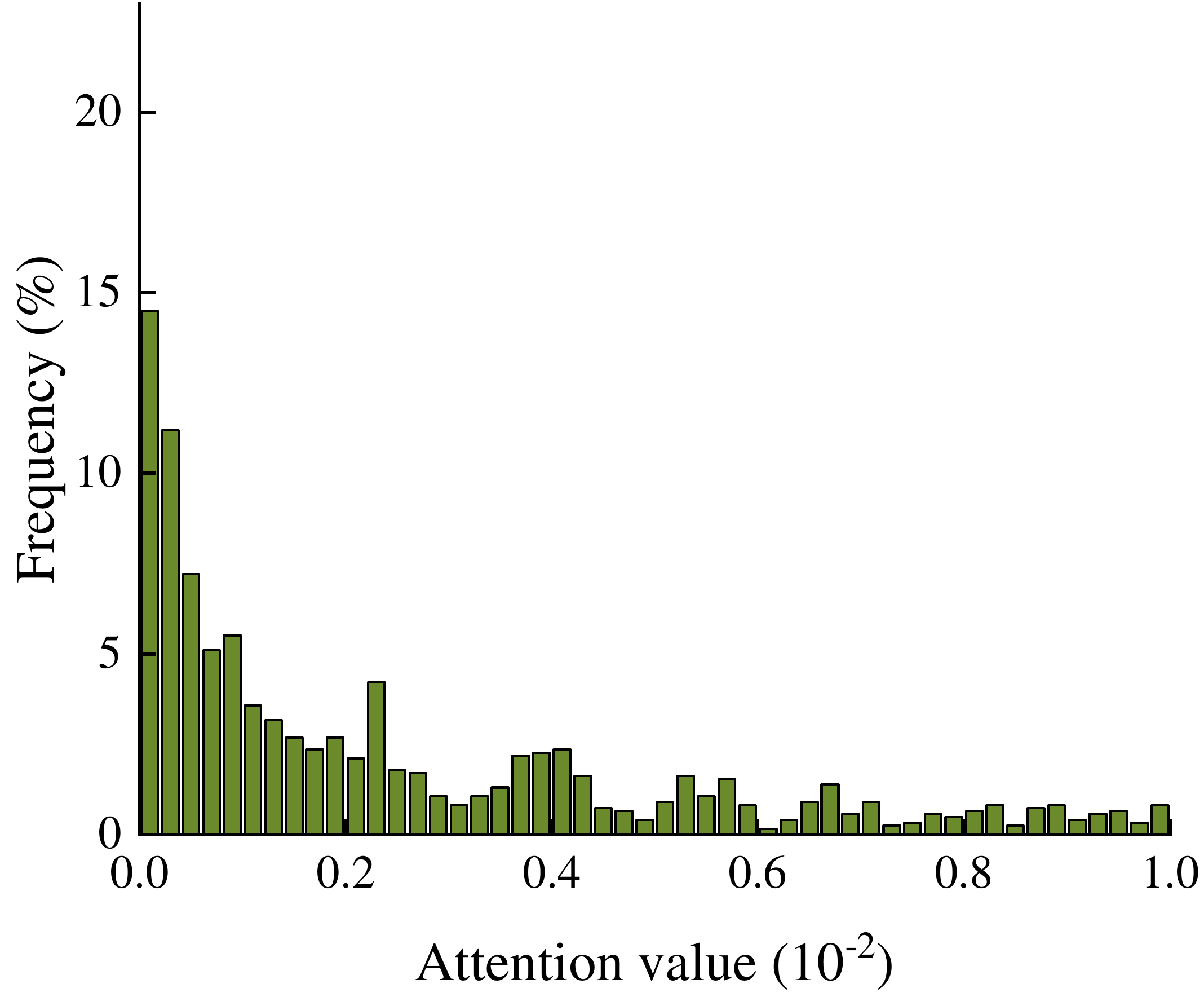}
  \label{fig:API_dimension2_sub1}
  }
  \quad
  \subfigure[Family Airpush]{
  \includegraphics[width=0.26\textwidth]{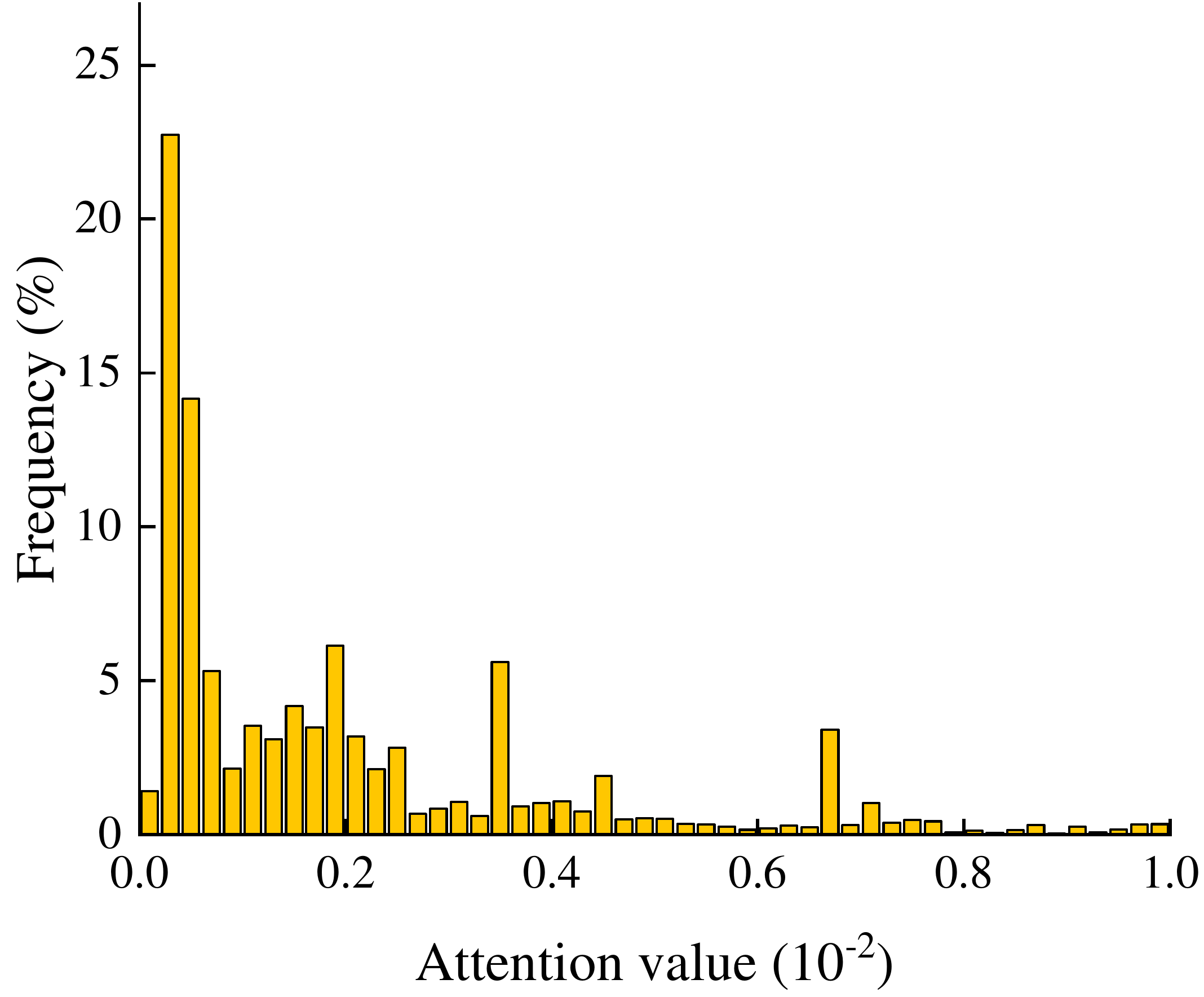}
  \label{fig:API_dimension2_sub2}
  }
  \quad
  \subfigure[Family BankBot]{
  \includegraphics[width=0.26\textwidth]{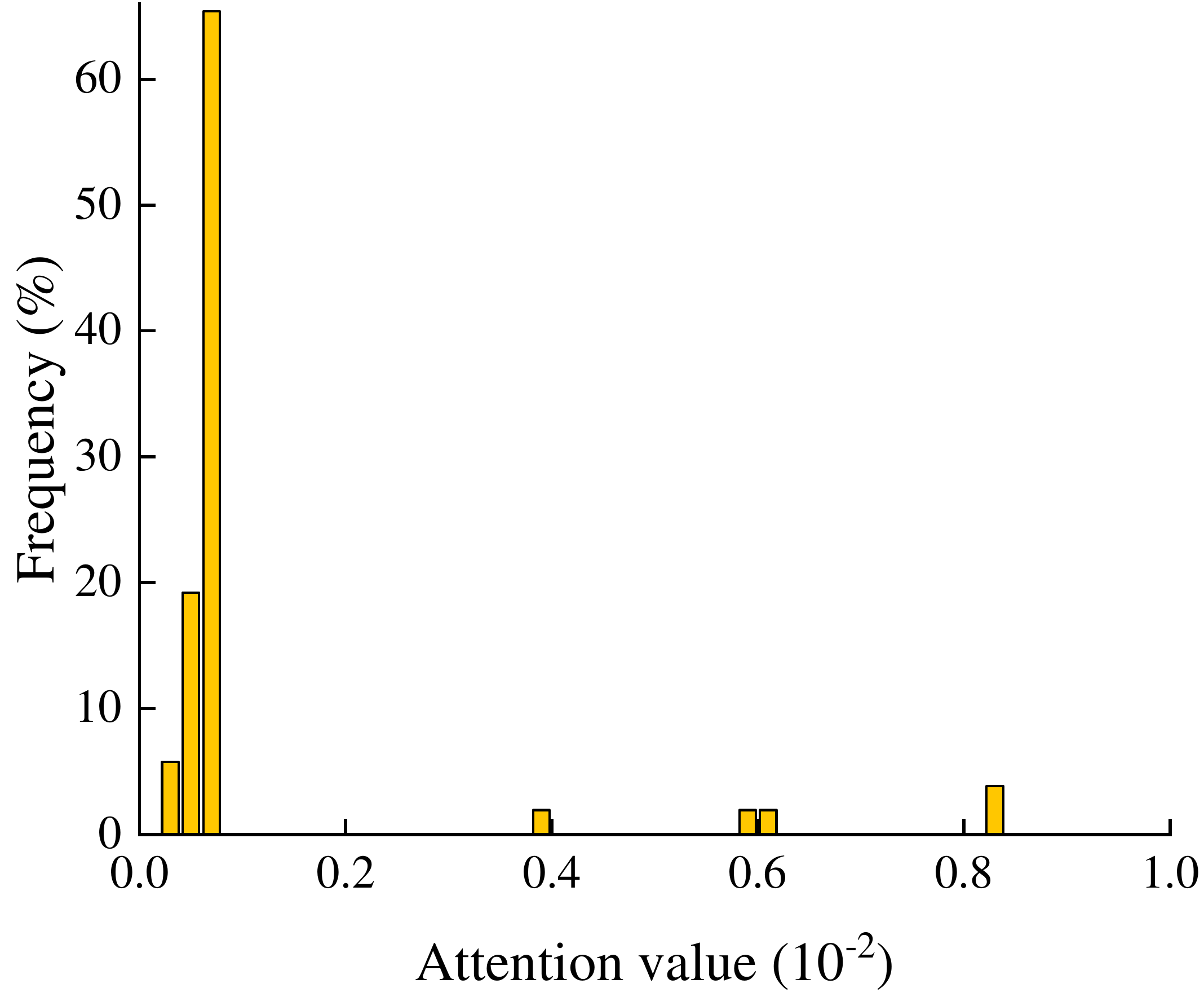}
  \label{fig:API_dimension2_sub3}
  }
  \quad
  \subfigure[Family Dowgin]{
  \includegraphics[width=0.26\textwidth]{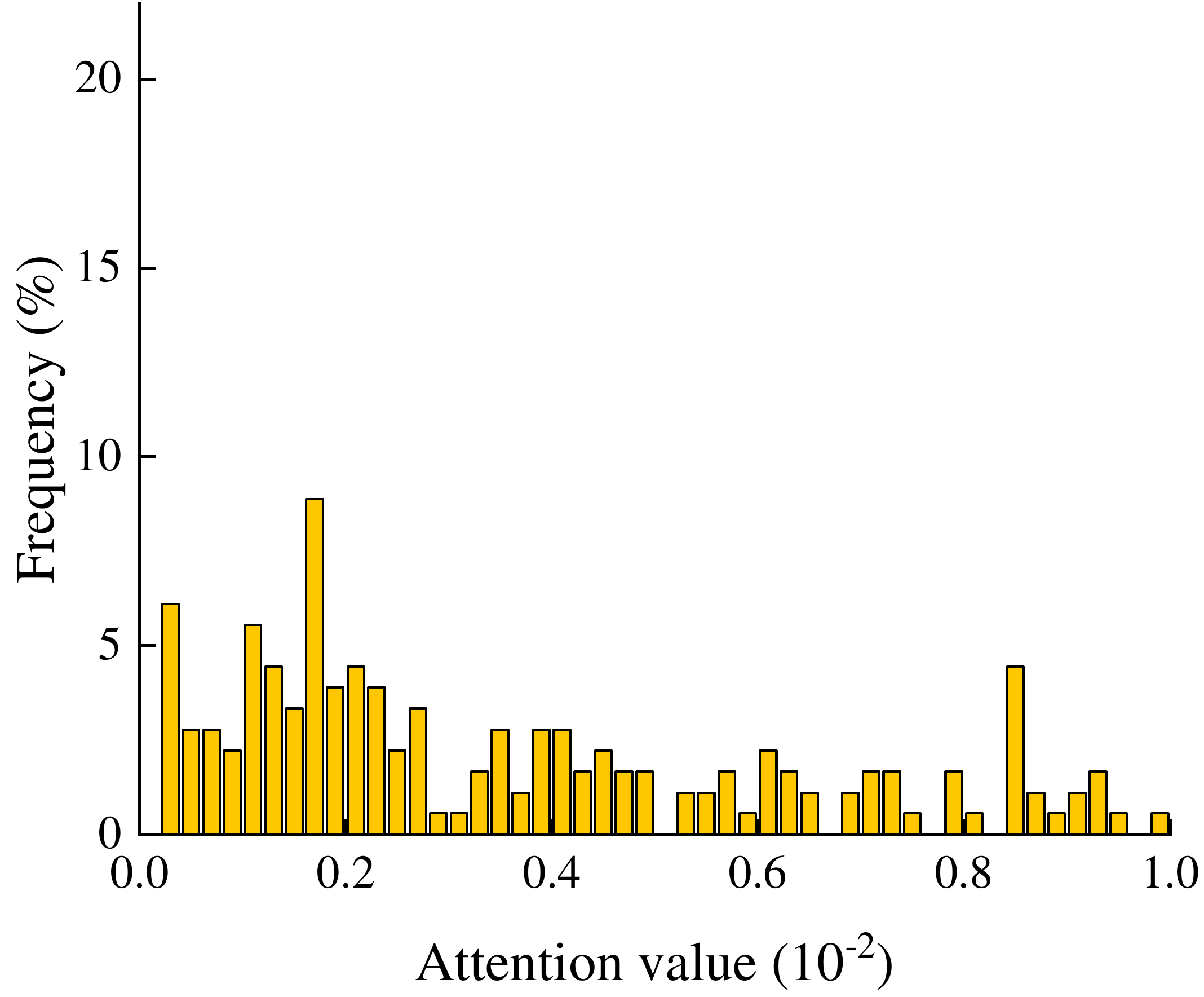}
  \label{fig:API_dimension2_sub4}
  }
  \quad
  \subfigure[Family Fusob]{
  \includegraphics[width=0.26\textwidth]{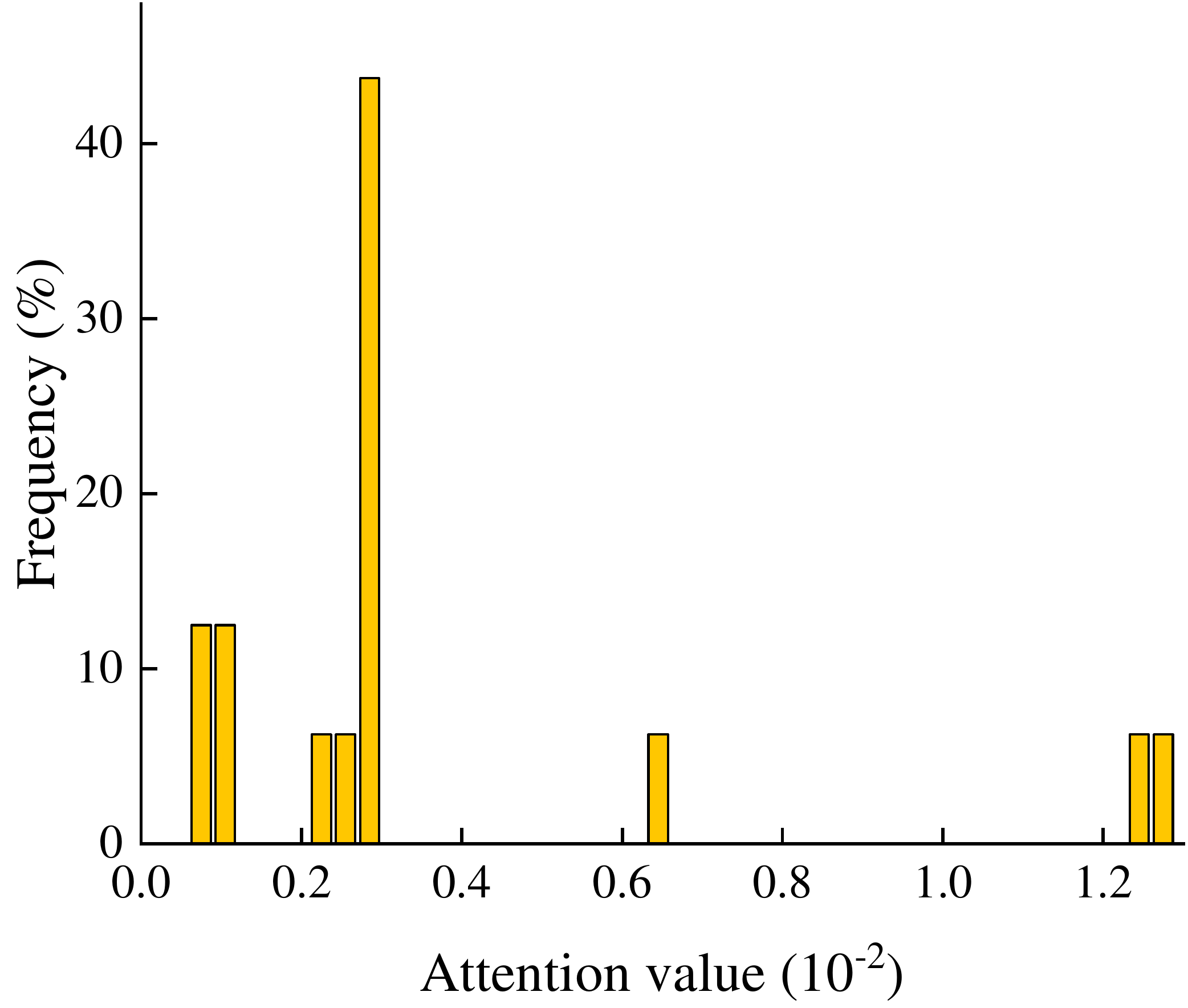}
  \label{fig:API_dimension2_sub5}
  }
  \quad
  \subfigure[Family FakeInst]{
  \includegraphics[width=0.26\textwidth]{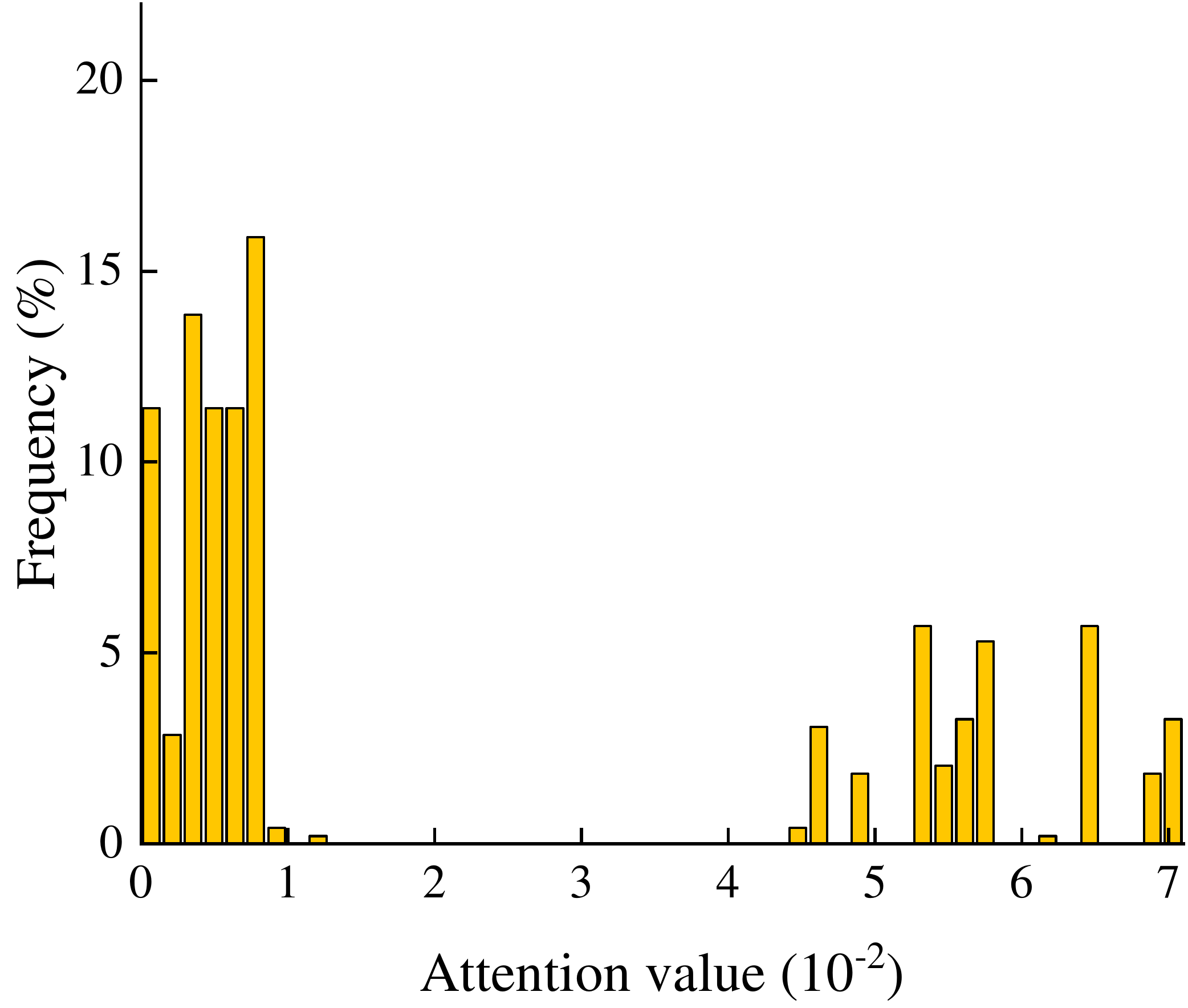}
  \label{fig:API_dimension2_sub6}
  }
  \quad
  \caption{API attention distribution of \textit{android/net/wifi/WifiInfo; getMacAddress()}}
  \label{fig:API_distribution2}
\end{figure*}

\begin{figure}[htbp]
  \centering
  \includegraphics[width=0.4\textwidth]{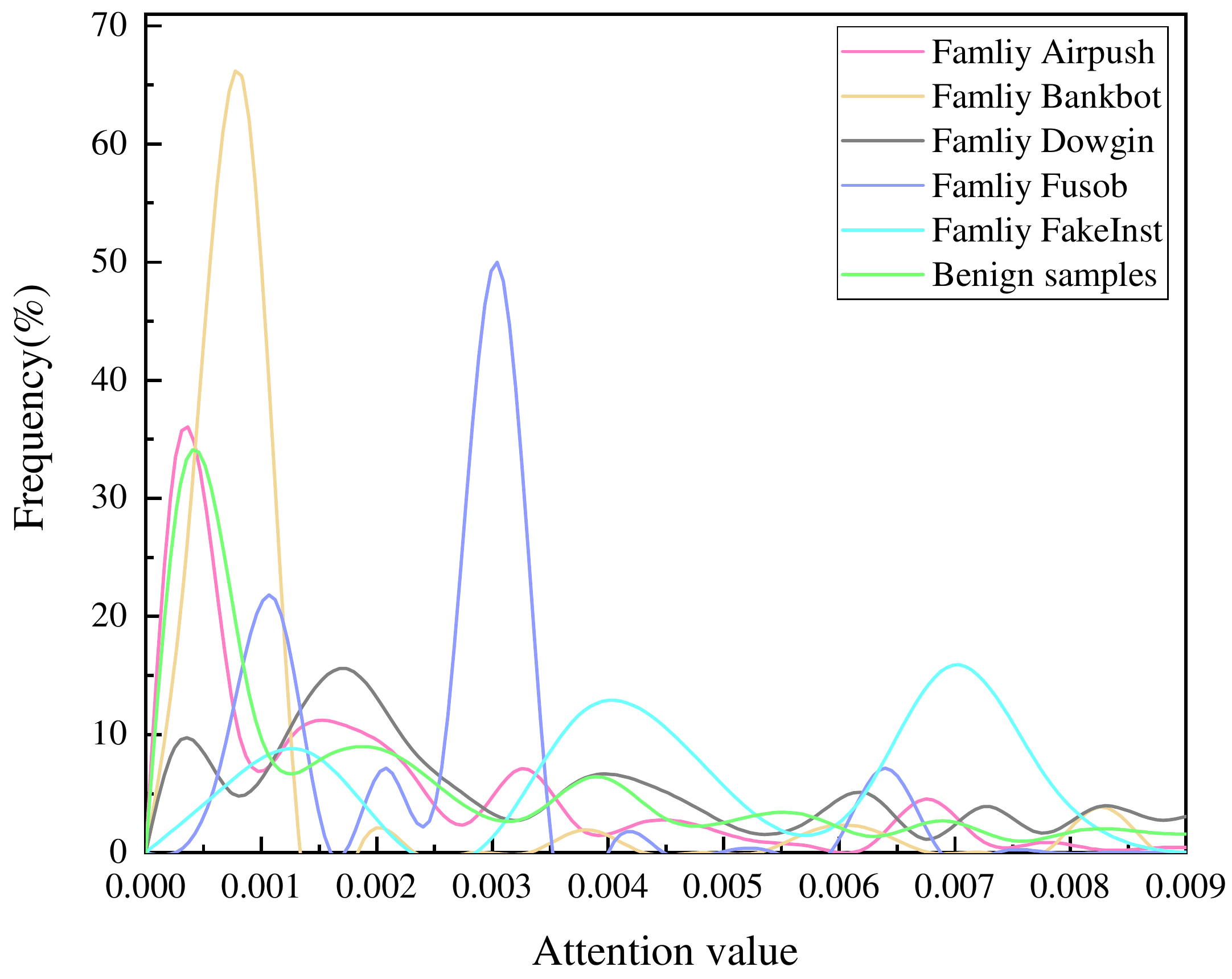}
  \caption{API attention distribution curves of \textit{android/net/wifi/ WifiInfo;getMacAddress()}}
  \label{fig:API_distribution3}
\end{figure}

Unlike basis APIs, the APIs that sensitive behavior may involve are treated differently in Droidetec. In Fig.\ref{fig:API_distribution2}, we take the API \textit{android/net/wifi/WifiInfo;getMacAddress()} (abbreviated as \textit{getMacAddress()}) as an example. Fig.\ref{fig:API_dimension2_sub1} details the frequency of attention values in benign samples while Figs. \ref{fig:API_dimension2_sub2} to \ref{fig:API_dimension2_sub6} are the case in 5 malware families.
In this case, Droidetec reflects the differences in attention of \textit{getMacAddress()}.
In benign samples, most of the attention values are less than 2$\times{\text{10}}^{\text{-3}}$, while in family \textit{FakeInst} 30\% of attention values exceed 4.43$\times{\text{10}}^{\text{-2}}$.
To make it more distinct, Fig. \ref{fig:API_distribution3} depicts the frequency distribution curves of attention value in interval (0, 9$\times{\text{10}}^{\text{-3}}$] for benign samples and 5 malware families.
The distribution curves of benign samples and family \textit{Airpush} are very close and are mostly in areas with smaller attention values, while attention in other families tends to be higher.
In addition, from family \textit{Airpush} to \textit{FakeInst}, the peak interval of frequency shifts to higher attention values, which reveals the semantic difference in malware families. The details are discussed in the following subsection.

In average, the API attention value of \textit{getMacAddress()} (5.542$\times{\text{10}}^{\text{-3}}$) is far more than that of \textit{onCreate()} (1.09$\times{\text{10}}^{\text{-4}}$). Generally, Droidetec concerns more about behavior-related APIs and places different emphasis on benign samples and different families.

\subsubsection{Semantic difference in malware families}

\begin{table*}
  \centering
  \caption{5-max APIs in malware families}
  \label{tab:5-max_API_family}
  \begin{tabular}{l|lll}
    \hline
    \textbf{Family}& \textbf{5-max API}& \textbf{Suspected rate}& \textbf{Average weight}
    \\
    \hline
    \multirow{5}{*}{Airpush}& android/location/Criteria;setAccuracy& 64.3\%& 0.073
    \\
    \cline{2-4}
    & android/location/Criteria;setCostAllowed& 60.2\%& 0.018\\
    \cline{2-4}
    & android/net/wifi/WifiInfo;getMacAddress& 60.0\%& 0.020\\
    \cline{2-4}
    & android/provider/Settings\$Secure;getString& 26.9\%& 0.013\\
    \cline{2-4}
    & android/telephony/TelephonyManager;getDeviceId& 25.5\%& 0.019\\
    \hline

    \multirow{5}{*}{BankBot}& android/telephony/TelephonyManager;getLine1Number& 54.7\%& 0.026
    \\
    \cline{2-4}
    & android/telephony/TelephonyManager;getDeviceId& 44.8\%& 0.034\\
    \cline{2-4}
    & android/telephony/SmsManager;getDefault& 42.2\%& 0.204\\
    \cline{2-4}
    & android/telephony/SmsMessage;createFromPdu& 38.4\%& 0.047\\
    \cline{2-4}
    & android/app/admin/DeviceAdminReceiver;onEnabled& 36.9\%& 0.013\\
    \hline

    \multirow{5}{*}{Dowgin}& android/telephony/TelephonyManager;getDeviceId& 57.2\%& 0.028
    \\
    \cline{2-4}
    & android/net/wifi/WifiInfo;getMacAddress& 45.5\%& 0.035\\
    \cline{2-4}
    & android/view/Display;getMetrics& 44.3\%& 0.025\\
    \cline{2-4}
    & android/telephony/TelephonyManager;getLine1Number& 35.1\%& 0.026\\
    \cline{2-4}
    & android/content/Context;getClassLoader& 22.1\%& 0.011\\
    \hline

    \multirow{5}{*}{DroidKungFu}& android/telephony/TelephonyManager;getDeviceId& 86.3\%& 0.029
    \\
    \cline{2-4}
    & android/telephony/TelephonyManager;getLine1Number& 65.0\%& 0.035\\
    \cline{2-4}
    & android/widget/RelativeLayout;onTrackballEvent& 56.7\%& 0.199\\
    \cline{2-4}
    & android/widget/RelativeLayout;setPressed& 54.3\%& 0.028\\
    \cline{2-4}
    & android/net/NetworkInfo;getExtraInfo& 48.3\%& 0.061\\
    \hline

    \multirow{5}{*}{FakeInst}& android/telephony/SmsManager;getDefault& 82.4\%& 0.082
    \\
    \cline{2-4}
    & android/telephony/SmsManager;sendTextMessage& 82.4\%& 0.048\\
    \cline{2-4}
    & android/telephony/SmsMessage;createFromPdu& 45.3\%& 0.028\\
    \cline{2-4}
    & android/telephony/TelephonyManager;getLine1Number& 44.5\%& 0.047\\
    \cline{2-4}
    & android/content/SharedPreferences\$Editor;commit& 36.1\%& 0.003\\
    \hline
  \end{tabular}
\end{table*}

We further test Droidetec's semantic analysis capabilities in different families.
Since the suspected methods in different malware vary from each other, we employed a unified expression for evaluation. As defined in Section\ref{semantic}, the $k$-suspect APIs are $k$ APIs with the largest attention values in a certain program. To describe a set of programs, we figure out the $n$ most frequency $k$-suspect APIs as $n$-max APIs. The $n$-max APIs are good indicators of a program set especially with similar attributes.


In Tab.\ref{tab:5-max_API_family}, we make statistics of $5$-max API by malware families with $k=200$.
APIs \textit{getLine1Number} and \textit{getDeviceId} of class \textit{android/telephony/TelephonyManager} tend to be treated with more emphasis, no matter which malware family they are from. These two API calls respectively attempt to obtain the IMEI (International Mobile Equipment Identity) code and the local number of the mobile phone.

Besides, each family has its behavioral characteristics. Airpush is a malware family that aggressively pushes advertising content to the device's notification bar. In the case of the Airpush family, Droidetec focuses on \textit{setAccuracy} and \textit{setCostAllowed}, which request Location Provider to provide location and direction information.

Family BankBot is a banking trojan that embodies stealing SMS, money transferring, GPS location tracking and so on. FakeInst malware appears to be installers for normal applications but sends SMS messages to premium-rate numbers or services when executed. In both families, 4 of the 5-max APIs from package \textit{android/telephony}, and the average attention weight of \textit{android/telephony/SmsManager;getDefault()} even reaches 0.204 in BankBot. These APIs are relative to mobile phone identification and SMS sending, which indicates the main behavioral characteristics of malware families.

In all, Droidetec changes its focus for different API sequences. Moreover, it also shows the distinction in semantic analysis when deals with various malware families.

\subsection{Automatic malicious code localization}
Based on the detection result and semantic attention, Droidetec eventually generates an analysis report if the program under test is detected as malware.

\subsubsection{An instance of malicious code localization}
\begin{figure}[htbp]
  \centering
    \includegraphics[width=\linewidth]{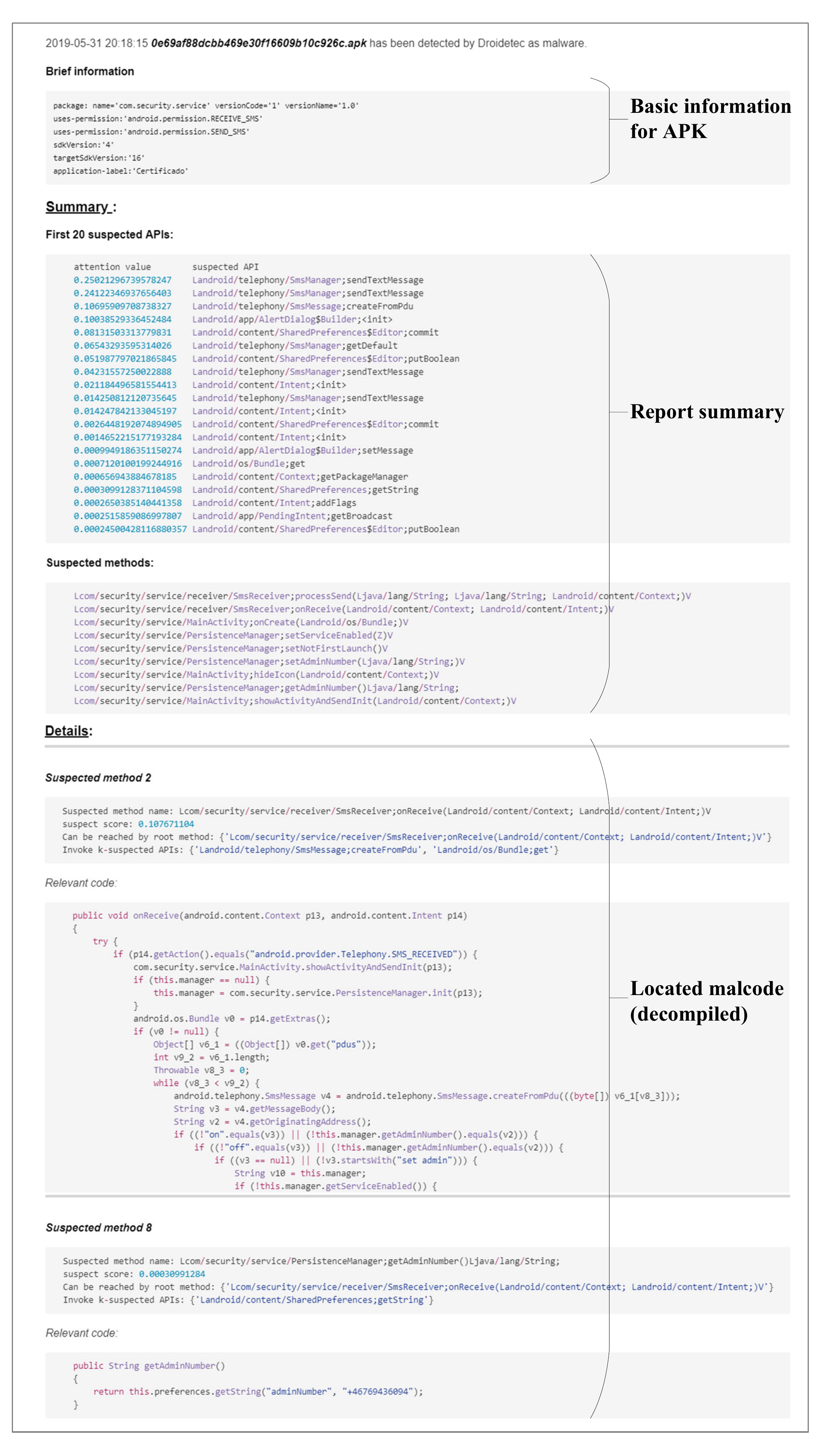}
    \caption{A report instance}
    \label{fig:report}
\end{figure}

We first take a sample from malware family \textit{Zitmo} as an instance to illustrate the analysis report. As shown in Fig.\ref{fig:report}, the analysis report contains 3 parts, the brief information, the summary and the details.

The first part offers basic information for this APK, including the package name, the requested permissions, the version of SDK (Software Development Kit) and so on.
The summary explains the first $k$ suspected APIs along with their attention values and the suspected methods, which indicate the corresponding package and class, the parameter type and the return type. In this instance, several of the top 20 suspected APIs appear repeatedly. For example, the API \textit{sendTextMessage()} appears 4 times with discrepant attention values, for it is invoked in different segments of behavior sequence.
Specific information about suspected methods is offered in the detail part, including the suspected socre (the weighted sum $sus(m)$ in Eq.\ref{eq:suspicious degree1}), the possible entry points, the invoked $k$-suspected APIs and the decompiled code.
Fig.\ref{fig:report} selects the suspected method 2 (\textit{onReceive}) and 8 (\textit{getAdminNumber}).
The main malicious point occurs in \textit{onReceive}, which monitors users' text messages. The communication with a specific number leaks sensitive information, including specific message content and source numbers. The program then receives instructions sent by the specific number and perform malicious acts. In the source code of \textit{onReceive}, the method \textit{getAdminNumber} is invoked. This method is also captured by Droidetec as a suspected method where the specific number ``$+46769436094$'' is exposed.

It should be mentioned that Droidetec can completely display malicious source code segments no matter they are obfuscated code or not. For the convenience of explanation, we select this malware sample with more straightforward class and method names.

\subsubsection{Localization evaluation}
As described in Sec.\ref{semantic}, top $n$ methods with the highest $sus(m)$ scores are selected as possibly malicious methods. A larger $n$ can cover more potentially malicious methods, but can also bring a mass of inaccurate localization. To determine an appropriate $n$ value and evaluate the localization effectiveness, we define the hit rate and accuracy that are respectively given by
\begin{equation}
  \text{hit rate} = \frac{N_{\text{hit}}}{N},
\end{equation}
\begin{equation}
  \text{accuracy} = \frac{\sum_{i=1}^{N}n^{'}_{i}}{N\times n}.
  \label{eq:accuracy}
\end{equation}
Droidetec is regarded to hit the malware, if one or more malicious code segments are successfully located. In total $N$ malicious samples, $N_{hit}$ pieces of malware are hit. In sample $i$, Droidetec correctly grabs $n^{'}_{i}$ true malicious methods among $n$ suspected methods, and Eq.\ref{eq:accuracy} utilizes the accuracy to measure the localization quality and limit the size of $n$.

\begin{figure}[htbp]
  \centering
    \includegraphics[width=0.75\linewidth]{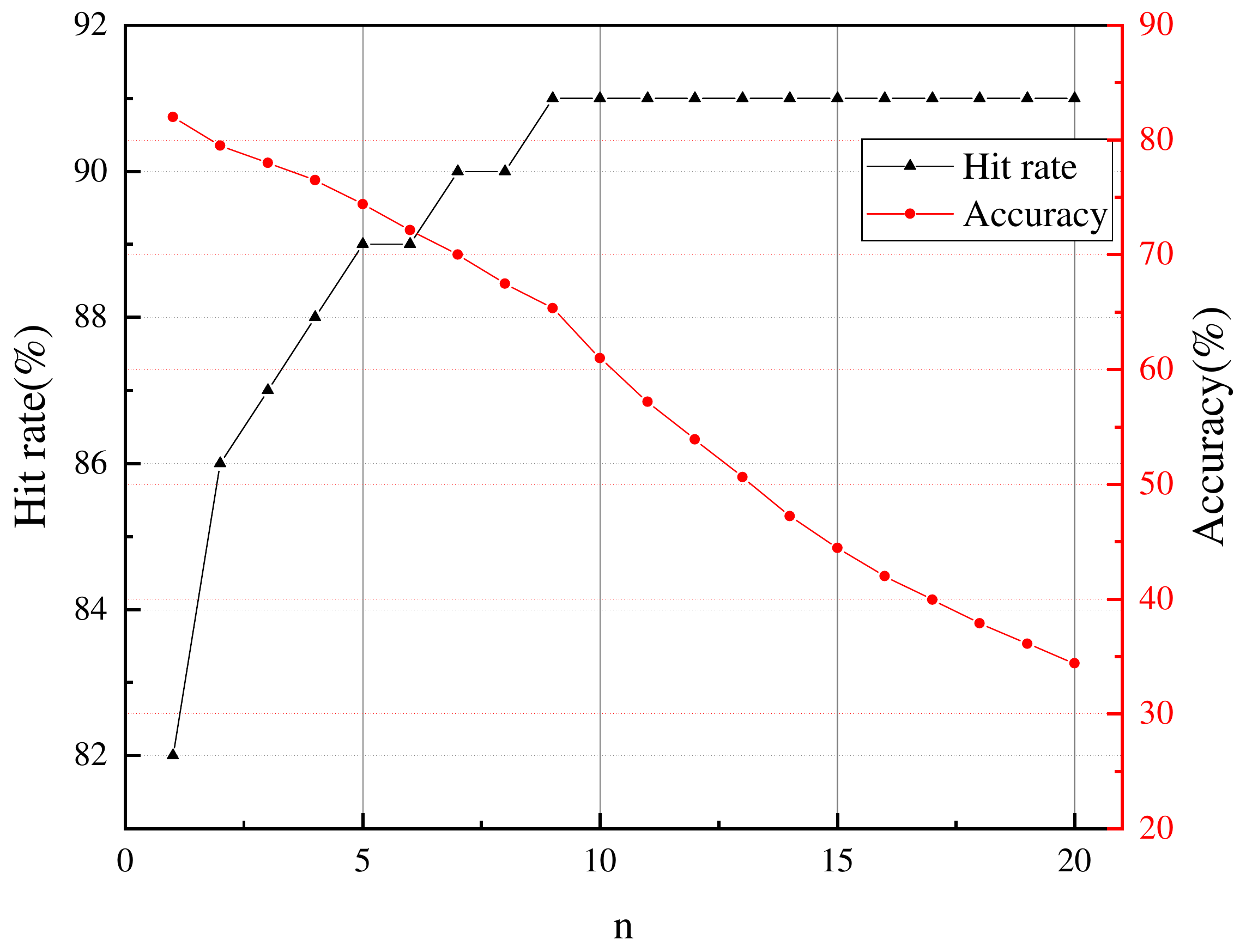}
    \caption{Hit rate and accuracy}
    \label{fig:locating_hit_accuracy}
\end{figure}

\begin{figure}[htbp]
  \centering
    \includegraphics[width=0.6\linewidth]{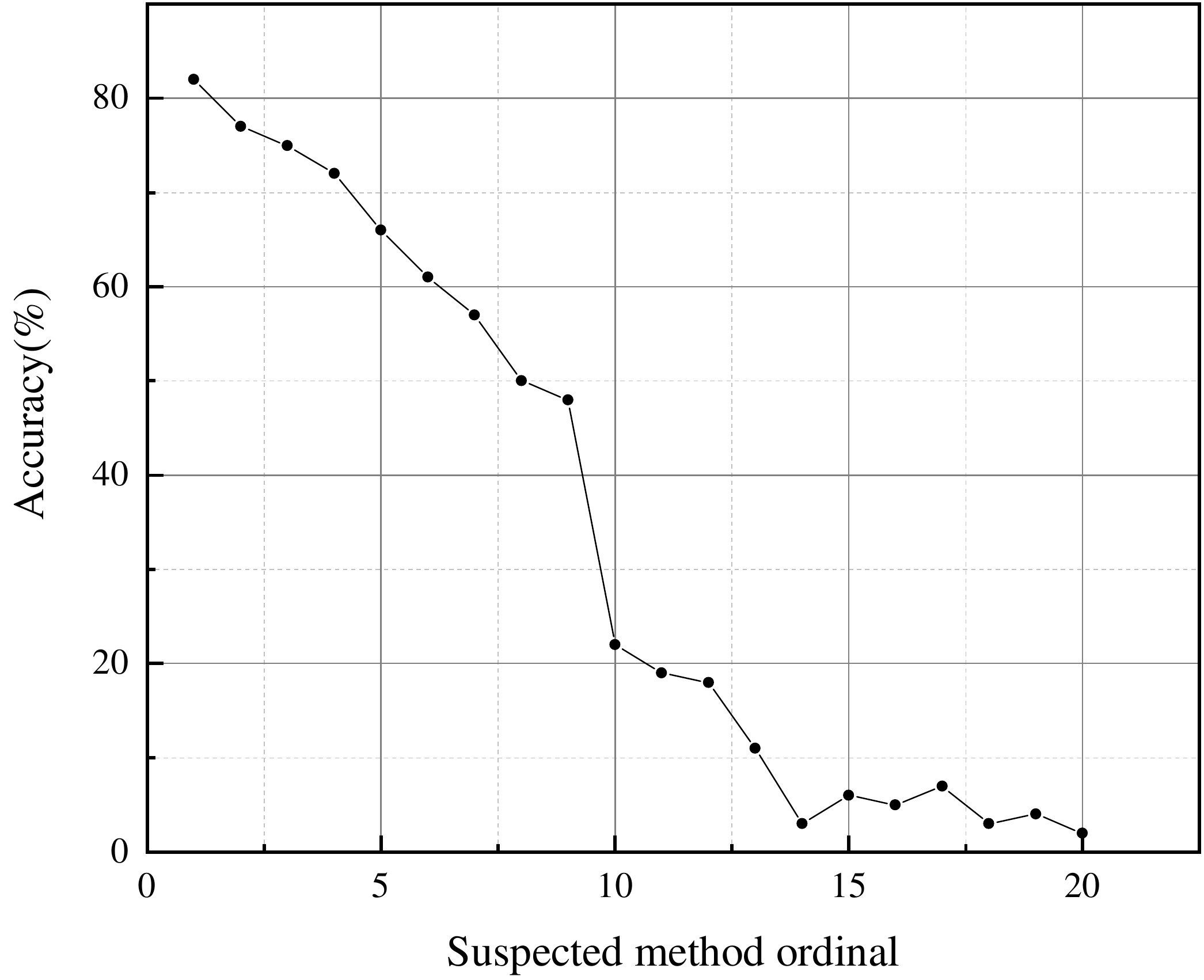}
    \caption{Accuracy of single methods}
    \label{fig:locating_single_accuracy}
\end{figure}

In Fig.\ref{fig:locating_hit_accuracy}, we manually inspected 100 pieces of malware and depict how hit rate and accuracy change with $n$.
Only by results of the first suspected methods ($n = 1$), we can achieve 82\% hit rate and accuracy. As $n$ increases, hit rate turns to be higher while accuracy decreases. It is comprehensible that the more suspected methods we select, the higher the possibility of catching malicious code becomes. The accuracy decreases, for suspected methods with lower $sus(m)$ are more possible to be erroneous judgements. Given that $n = 1$, and the $m^{\text{th}}$ suspected method is the only localization result, the accuracy in Eq.\ref{eq:accuracy} turns into the single accuracy of method $m$. Fig.\ref{fig:locating_single_accuracy} shows the single accuracy from the $1^{\text{st}}$ suspected method to the $20^{\text{th}}$, and there is a sharp decline after the $9^\text{th}$ method, which results in the corresponding decline of overall accuracy in fig.\ref{fig:locating_hit_accuracy}. Meanwhile, the hit rate no longer increases when $n>9$. In this case, Droidetec offers 9 most suspected methods by default and achieves 91\% hit rate with 65.3\% overall accuracy.

\subsection{Limitation}
Droidetec implements malware detection by analyzing the Dalvik opcode of a program. Hence, libraries linked during running-time such as .so files do not belong to our analysis scope. It is not the case in practice as Droidetec can be easily combined with other approaches specific to native shared libraries detection.

\begin{table}[htbp]
  \centering
    \caption{Droidetec time consumption of malware $< 10M$ in size}
    \label{tab:time_consumption}
    \begin{tabular}{c|c|c|c}
      \hline
      \multicolumn{4}{c}{\textbf{Time consumption}}\\
      \hline
      \textbf{Feature extraction} & \textbf{Classification} & \textbf{Report generation} & \textbf{Overall}\\
      \hline
      1.263s & 0.31s & 0.109s & 1.682s\\
      \hline
    \end{tabular}
    \\[0.1in]
    \begin{tabular}{c|c|c}
      \hline
      \multicolumn{3}{c}{\textbf{Feature extraction}}\\
      \hline
      \textbf{Deep invocation traversal} & \textbf{Sequence generation} & \textbf{Vectorization}\\
      \hline
      0.89s & 0.033s & 0.34s\\
      \hline
    \end{tabular}

\end{table}

A limitation of Droidetec is time consumption.
Tab.\ref{tab:time_consumption} indicates the time consumption in all stages of detection when the malware samples are less than $10M$ in size.
It takes Avira $0.54s$ per program on average, while Droidetec spends $1.682s$ per program to analyse the same samples on the same device.
It is evident that sequence feature extraction takes up most of the time (75\%). As stated in Section \ref{introduction}, we aim to propose a solution that implements a deeper and more efficient static detection to substitute for the dynamic. Therefore, we have to take an effort to work out the calling relationship and behavior sequence of a program.
Theoretically, we use
\begin{equation}
T(n)=(n_{avg})^d, \quad(0<n_{avg}<<n)
\label{eq:time_complexity}
\end{equation}
to measure the time complexity of the feature extraction process.
In Eq. \ref{eq:time_complexity}, $n$ represents the method amount of a program, $n_{avg}$ is the average number of times a method calls other methods and $d$ is the average invocation depth.
Although $b_{avg}$ is far less than $n$, the time complexity is higher than the $O(n)$ of the common extraction methods\cite{yuan2014droid, arp2014drebin, 8629067, 8406618} which only require a direct code walk.
We have optimized the time overhead with dynamic planning, which adopts extra storage space to make the complexity as close as possible to $O(n)$.
While in the worst case ($d$ is large), the time spend during the deep invocation traversal increases greatly.

\section{Related Work}\label{sec4}
In the past few years, research and experimentation of Android malware detection methods have been continually evolving. According to the categorization and analyses of Arshad et al.\cite{arshad2016android}, we can have an overall view of the current Android malware landscape and corresponding detection methods.
\subsection{Android malware detection}
A general detection for Android malware is the permission-based method\cite{shin2009towards, 8404302, arora2019permpair}, which analyses the manifest and notifies about the over-privileged applications. These methods, though fast, are difficult to achieve a guaranteed accuracy in practice.
Contemporarily, the classification by simple permissions is not appropriate as most Android applications tend to be functional complexity and request more permissions.

Signature-based approaches\cite{zheng2013droid, Faruki:2013:ARS:2523514.2523539, saracino2016madam, rehman2018machine} extract features to create a unique signature for each application. The program under test will evaluate to malicious if its signature matches with existing malware families'.
For example,
AndroSimilar\cite{Faruki:2013:ARS:2523514.2523539} generates the variable length signature for the application under test. It measures syntactic file similarity of the whole file instead of just opcodes for faster detection and implements classification based on similarity percentage.
Madam\cite{saracino2016madam} implements a signature-based approach that considers behavioral patterns from known malware misbehaviors.
Although several improvements have been proposed, the signature-based detection performs unsatisfactorily in dealing with unknown malware. In most cases, it can be relatively easy for malware to evade this kind of detection by adding simple obfuscation methods.

Dynamic analysis\cite{burguera2011crowdroid, enck2014taintdroid, Wong2016IntelliDroidAT, feng2018novel} examines the application during execution.
Crowdroid\cite{burguera2011crowdroid} collects the system calls during program running time. A clustering algorithm is adopted in identifying malware and normal programs.
TaintDroid proposed by Enck et al.\cite{enck2014taintdroid} is a famous dynamic taint tracking framework that labels and tracks sensitive data during a source-sink period. It performs well in information flow tracking and can effectively avoid privacy leaking.
IntelliDroid\cite{Wong2016IntelliDroidAT} provides a generic Android input generator that can produce inputs specific to a dynamic analysis tool. It claimed that only a small number of inputs and a small part of the program execution are needed.
Dynamic malware detection tends to be designed for several certain malicious behaviors and is time-consuming as the tested application has to keep running for a long time until anomalies occur.
Dynamic detection can be easily blocked in places where human operations are necessary. Thus, in most cases, only manual dynamic detection can be satisfactory. Droidetec can be combined directly with these dynamic methods. Using Droidetec as preprocessing can greatly reduce the workload of manual analysis.

\subsection{Combination with machine learning}
The development of machine learning has opened up a new way to malware detection, where machine learning algorithms, especially several deep learning algorithms, have been applied to feature process and classification\cite{arp2014drebin, yuan2014droid, hou2016deep4maldroid, mclaughlin2017deep, li2018significant, 8629067, 8443370, chen2019android}.

For instance, Drebin \cite{arp2014drebin} implements an effective and explainable detection for Android malware that extracts 8 feature sets from the manifest and disassembled code. It utilizes linear SVM (Support Vector Machines) for this task.
Droid-Sec was proposed by Yuan et al.\cite{yuan2014droid}, which extracts 202 features including required
permission, sensitive API and dynamic behavior. The comparison with traditional models such as SVM and C4.5 etc. demonstrates that the DBN (deep belief network) they adopt has the best performance.
Mclaughlin et al.\cite{mclaughlin2017deep} proposed a deep CNN (Convolutional Neural Network) based detection system that extracts the raw opcode sequence from a disassembled program as features. However it only analyzes 218 defined opcodes, most of which are not behavior related. Invoking related instructions with different operands such as \textit{android/os/SystemClock;sleep()}, \textit{android/telephony/TelephonyManager;getSimOperatorName()} and \textit{android/net/NetworkInfo;getDetailed()} are regarded as the same feature, which is not reasonable enough.

\subsection{Malicious code localization}
There have been two known studies on Android malicious code localization, which are both based on the CFG. Li et al.\cite{Li2017} focused on finding the hooks between carrier and rider code, and defined two hook types which differ in the way rider code is triggered: through method calls or the Android event system. In our case, $Type_{1,2}$ are within our detection scope, since the concept of root method encompasses both types.
Narayanan et al.\cite{Narayanan2018} analyzed the Inter-procedural CFG and assigned an m-score to quantify the statistical significance of malice operations. The limitation of the two graph-based methods is that the analysis accuracy will be affected by incorporating benign subgraph features, while the forget gate in LSTM helps Droidetec mitigate this negative impact.

Overall, previous machine learning based solutions use extensive features and only care about achieving outstanding classification results. In our case, Droidetec provides accurate detection along with the retracing to suspected segments.

\section{Conclusion}\label{sec5}
This paper presents Droidetec, a static and automatic analysis framework for Android malware detection using a deep neural network based approach. The feature extraction method is utilized to traverse all the invocation processes in an orderly manner. Besides, Droidetec goes beyond others in that it provides automatic analysis which indicates the suspected code segments. That means, to a certain extent, program analysts could free themselves from reading complex, obfuscated malicious code, and efficiently discover the malicious patterns.
In the future, a more accurate method for multi-classification of various malware families will be the focus of our work.
Besides, we are looking for a new way to improve the malicious code localization in that we will design the range of code segments instead of reporting the complete decompiled methods.

\bibliographystyle{IEEEtran}
\bibliography{reference}
\ifCLASSOPTIONcaptionsoff
  \newpage
\fi


\end{document}